\def\year{2018}\relax
\documentclass[letterpaper]{article} %DO NOT CHANGE THIS
                      
\usepackage{aaai18}  %Required
\usepackage{times}  %Required
\usepackage{helvet}  %Required
\usepackage{courier}  %Required
\usepackage{url}  %Required
\usepackage{graphicx}  %Required
\usepackage{amssymb}
\usepackage{amsmath}
\usepackage{mathtools}
\usepackage{amsfonts}
\usepackage{bm} 
\usepackage{graphicx}
\usepackage{rotating}
\usepackage{tikz}
\usetikzlibrary{shapes}
\usetikzlibrary{fit}
\usetikzlibrary{chains}
\usetikzlibrary{arrows}

\tikzstyle{latent} = [circle,fill=white,draw=black,inner sep=1pt,
minimum size=22pt, font=\fontsize{12}{10}\selectfont, node distance=1]
\tikzstyle{obs} = [latent,fill=gray!25]
\tikzstyle{const} = [rectangle, inner sep=0pt, node distance=1]
\tikzstyle{factor} = [rectangle, fill=black,minimum size=5pt, inner
sep=0pt, node distance=0.4]
\tikzstyle{det} = [latent, diamond]

\tikzstyle{plate} = [draw, rectangle, rounded corners, fit=#1]
\tikzstyle{wrap} = [inner sep=2pt, fit=#1]

\tikzstyle{newwrap} = [inner sep=0pt, fit=#1]
\tikzstyle{gate} = [draw, rectangle, dashed, fit=#1]

\tikzstyle{caption} = [font=\footnotesize, node distance=0] %
\tikzstyle{plate caption} = [caption, node distance=0, inner sep=0pt,
above right=0pt and 0pt of #1.north east] %
\tikzstyle{new plate caption} = [caption, node distance=0, inner sep=0pt,
above right=0pt and 0pt of #1.south west] %
\tikzstyle{factor caption} = [caption] %
\tikzstyle{every label} += [caption] %

\tikzset{>={triangle 45}}

\newcommand{\edge}[3][]{ %
  \foreach \x in {#2} { %
    \foreach \y in {#3} { %
      \path (\x) edge [->,#1] (\y) ;%
    } ;
  } ;
}

\newcommand{\plate}[4][]{ %
  \node[newwrap=#3] (#2-newwrap) {}; %
  \node[plate caption=#2-newwrap] (#2-caption) {#4}; %
  \node[plate=(#2-newwrap)(#2-caption), #1] (#2) {}; %
}

\usepackage{tabularx,hhline}
\usepackage{array}
\usepackage{tabulary}
\newcolumntype{K}[1]{>{\centering\arraybackslash}p{#1}}
\usepackage{bbm}
\usepackage{array}
\usepackage{multirow}
\usepackage{subcaption}
\usepackage{bbm}
\usepackage{tabularx,booktabs} 
\usepackage{standalone}
\usepackage[acronym,smallcaps,nowarn]{glossaries}
\usepackage{aaai18}  %Required
\usepackage{times}  %Required
\usepackage{helvet}  %Required
\usepackage{courier}  %Required
\usepackage{url}  %Required
\usepackage{graphicx}  %Required
\usepackage[algoruled]{algorithm2e}
\usepackage{listings}
\usepackage{fancyvrb}
\fvset{fontsize=\normalsize}

\usepackage[
  separate-uncertainty = true,
  multi-part-units = repeat
]{siunitx}

 \tikzstyle{legend_isps}=[rectangle, thin, 
                      , draw,minimum height = 0.6cm]

\begin{document}
\title{A Poisson Gamma Probabilistic Model for Latent Node-group Memberships in Dynamic Networks}
\author{Sikun Yang, Heinz Koeppl\\
Department of Electrical Engineering and Information Technology, Technische Universit\"at Darmstadt\\64283 Darmstadt, Germany\\
\{sikun.yang, heinz.koeppl\}@bcs.tu-darmstadt.de
}
\maketitle

%% ABSTRACT
\begin{abstract}
We present a probabilistic model for learning from dynamic relational data, wherein the observed interactions among networked nodes are modeled via the Bernoulli Poisson link function, and the underlying network structure are characterized by \emph{nonnegative} latent node-group memberships, which are assumed to be gamma distributed. 
The latent memberships evolve according to Markov processes.
The optimal number of latent groups can be determined by data itself. The computational complexity of our method scales with the number of non-zero links, which makes it scalable to large sparse dynamic relational data. We present batch and online Gibbs sampling algorithms to perform model inference. Finally, we demonstrate the model's performance on both synthetic and real-world datasets compared to state-of-the-art methods.
\end{abstract}

%% INTRODUCTION
\section{Introduction}
Considerable work has been done on the analysis of static networks in terms of community detection or link prediction~\cite{PHoff,IRM,MMSBM}. However, due to the temporal evolution of nodes (e.g. individuals), their role within a network can change and hence observed links among nodes may appear or disappear over time~\cite{Sci,SSN}. Given such dynamic network data, one may be interested in understanding the temporal evolution of groups in terms of their size and node-group memberships and in predicting missing or future unobserved links based on historical records.

A dynamic network of $N$ nodes can be represented as a sequence of adjacency matrices $\mathbf{b}^{(t)} \in \{0,1\}^{N\times N}$, $t = 1,\ldots,T$, where $b_{mn}^{(t)} = 1$ indicates the presence of a link between node $m$ and $n$ at time point $t$ and $b_{mn}^{(t)} = 0$ otherwise. For the sake of clarity we focus on undirected and unweighted networks but the presented method can be extended to weighted networks via the compound Poisson distribution~\cite{CPD} and to multi-relational networks~\cite{KnowBase,KnowGraph}.    

Many of the current probabilistic methods for dynamic networks map observed binary edge-variables (either links or non-links) to latent Gaussian random variables via the logistic or probit function~\cite{LGP}. The time-complexity of these approaches often scales quadratically with the number of nodes, i.e., $\mathcal{O}(N^2)$. This will become infeasible for large networks and will become especially inefficient for sparse networks. In this work we leverage the Bernoulli-Poisson link function~\cite{BLVM,EPM,FCEF} to map an observed binary edge to a latent Poisson count random variable, which leads to a computational cost that only scales with the number of non-zero edges. As large, real-world networks are usually very sparse, the proposed method yields significant speed-up and enables the analysis of larger dynamic networks. We allow for a time-varying nonnegative degree of membership of a node to a group. We realize this by constructing a gamma Markov chain to capture the time evolution of those latent membership variables. Inspired by recent advances in data augmentation and marginalization techniques~\cite{NBP}, we develop an easy-to-implement efficient Gibbs sampling algorithm for model inference. We also present an online Gibbs sampling algorithm that can process data in mini-batches and thus readily scales to massive sparse dynamic networks. The algorithms performs favorably on standard datasets when compared to state-of-the-art methods~\cite{DRIFT,DGPPF,EPM}.

%% MODEL
\section{Dynamic Poisson Gamma Membership Model}
In the proposed model, each node $n$ is characterized by a time-dependent latent membership variable $\phi_{nk}^{(t)}$ that determines its interactions or involvement in group $k$ at the $t\mbox{-}\mathrm{th}$ snapshot of the dynamic networks. This latent node-group membership is modeled by a gamma random variable and is, thus, naturally \emph{nonnegative real-valued}.  
This is contrast to \emph{multi-group memberships} models (or latent feature models)~\cite{DRIFT,LFP,MGMM} where each node-group membership is represented by a \emph{binary} latent feature vector.  These models assume that each node either associates to one group or not -- simply by a binary feature. The proposed model on the other hand can characterize how strongly each node associates with multiple groups.

\noindent\textbf{Dynamics of latent node-group memberships.} For dynamic networks, the latent node-group membership $\phi_{nk}^{(t)}$ can evolve over time to interpret the interaction dynamics among the nodes. For example, latent group $k$ could mean \textquotedblleft play soccer\textquotedblright\ and $\phi_{nk}^{(t)}$ could mean how frequently person $n$ plays soccer or how strongly person $n$ likes playing soccer. The person's degree of association to this group could be increasing over time due to, for instance, increased interaction with professional soccer players, or decreasing over time as a consequence of sickness. Hence, in order to model the temporal evolution of the latent node-group memberships, we assume the individual memberships to form a gamma Markov chain. More specifically, $\phi_{nk}^{(t)}$ is drawn from a gamma distribution, whose shape parameter is the latent membership at the previous time
\begin{align}
\phi_{nk}^{(t)} & \sim \mathrm{Gam}(\phi_{nk}^{(t-1)}/{\tau}, {1}/{\tau}),\qquad \text{for}\ t = 1,\ldots, T \notag\\
\phi_{nk}^{(0)} & \sim \mathrm{Gam}(g_0, 1/h_0), \notag
\end{align}
where the parameter $\tau$ controls the variance 
without affecting the mean, i.e.,  $\mathsf{E}[\phi_{nk}^{(t)} \mid\phi_{nk}^{(t-1)}, \tau] = \phi_{nk}^{(t-1)}$. 

\noindent\textbf{Model of latent groups.}
We characterize the interactions or correlations among latent groups by a matrix $\mathbf{\Lambda}$ of size $K\times K$, where $\lambda_{kk'}$ relates to the probability of there being a link between node $n$ affiliated to group $k$ and node $m$ affiliated to group $k'$.
Specifically, we assume the latent groups to be generated by the following hierarchical process: we first generate a separate weight for each group as
\begin{align}
r_k & \sim \mathrm{Gam}({\gamma_0}/{K}, {1}/{c_0}), \label{eq_pr_r}%\label{eq_pr_r}
\end{align}
and then generate the inter-group interaction weight $\lambda_{kk'}$ and intra-group weight $\lambda_{kk}$ as  
\begin{align}
 \lambda_{kk'} &\sim \begin{cases} \mathrm{Gam}(\xi r_{k}, {1}/{\beta}), & \text{if}\ k = k' \\ \mathrm{Gam}(r_{k}r_{k'}, {1}/{\beta}), & \text{otherwise}\end{cases} \label{eq_pr_lambda}
\end{align} 
where $\xi\in\mathrm{R}_{> 0}$ and $\beta\in\mathrm{R}_{> 0}$.
The reasonable number of latent groups can be inferred from dynamic relational data itself by the \emph{shrinkage} mechanism of our model. 
More specifically, for fixed $\gamma_0$, the redundant groups will effectively be shrunk as many of the groups weights tend to be small for increasing $K$. Thus, the interaction weights $\lambda_{kk'}$ between the redundant group $k$ and all the other groups $k'$, and all the node-memberships to group $k$ will be shrunk accordingly. 
In practice, the intra-group weight $\lambda_{kk}$ would tend to almost zero if $\lambda_{kk}\sim \mathrm{Gam}(r_k^2, 1/\beta)$ for small $r_k$, and the corresponding groups will disappear inevitably. Hence, we use a separate variable $\xi$ to avoid overly shrinking of small groups with less interactions with other groups.
As $\gamma_0$ has a large effect on the number of the latent groups, we do not treat it as a fixed parameter but place a gamma prior over it, i.e.,  $\gamma_0 \sim \mathrm{Gam}(1, 1)$.
 Given the latent node-group membership $\phi_{nk}^{(t)}$ and the interaction weights $\lambda_{kk'}^{(t)}$ among groups, the probability of there being a link between node $m$ and $n$ is given by
\begin{align}\label{link}
b_{mn}^{(t)} &\sim \mathrm{Bern}\left(1 - \exp\left\{-\sum_{k = 1}^{K}\sum_{k' = 1}^{K}\lambda_{kk'} \phi_{nk}^{(t)}\phi_{mk'}^{(t)}\right\}\right).
\end{align}
Interestingly, we can also generate $b_{mn}^{(t)}$ by truncating a latent count random variable $x_{mn}^{(t)}$ at $1$, where $x_{mn}^{(t)}$ can be seen as the integer-valued \emph{weight} for node $m$ and $n$ and can be interpreted as the number of times the two nodes interacted. More specifically, $b_{mn}^{(t)}$ can be drawn as 
\begin{align}
b_{mn}^{(t)} & = \mathbbm{1} (x_{mn}^{(t)} \geq 1),\\
x_{mn}^{(t)} & \sim \mathrm{Po}\left(\sum_{k = 1}^{K}\sum_{k' = 1}^{K}\lambda_{kk'} \phi_{nk}^{(t)}\phi_{mk'}^{(t)}\right), 
\end{align}
where $\mathrm{Po}$ indicates the Poisson distribution.  We can obtain Eq.~(\ref{link}) by marginalizing out the latent count $x_{mn}^{(t)}$ from the above expression. 
	The conditional distribution of the latent count $x_{mn}^{(t)}$ can then be written as 
\begin{align}
(x_{mn}^{(t)} \mid b_{mn}^{(t)}, \mathbf{\Phi, \Lambda})\sim b_{mn}^{(t)}\mathrm{Po}_{+}\left(\sum_{k = 1}^{K}\sum_{k' = 1}^{K}\lambda_{kk'} \phi_{nk}^{(t)}\phi_{mk'}^{(t)}\right), \notag
\end{align}
where $x\sim\mathrm{Po}_{+}(\sigma)$ is the zero-truncated Poisson distribution with probabilisty mass function (PMF) $f_X(x|\sigma) = (1-e^{-\sigma})^{-1}\sigma^{x}e^{-\sigma}/x!$ and $x\in\mathbb{Z}_{>0}$ and $\mathbf{\Phi}$ denotes the set of all node-group membership variables. The usefulness of this construction for $b_{nm}^{(t)}$ will become clear in the inference section. We note that the latent count $x_{mn}^{(t)}$ only needs to be sampled for $b_{mn}^{(t)}=1$, using rejection sampling detailed in~\cite{EPM}.   
To this end the proposed hierarchical generative model is as follows. 
\begin{align}
\phi_{nk}^{(t)} & \sim \mathrm{Gam}(\phi_{nk}^{(t-1)}/{\tau}, {1}/{\tau}),\ \ \text{for}\ t = 1,\ldots, T \notag\\
\phi_{nk}^{(0)} & \sim \mathrm{Gam}(g_0, 1/h_0),  \notag\\
r_k & \sim \mathrm{Gam}({\gamma_0}/{K}, {1}/{c_0}), \notag \\
 \lambda_{kk'} &\sim \begin{cases} \mathrm{Gam}(\xi r_{k}, {1}/{\beta}), & \text{if}\ k = k' \\ \mathrm{Gam}(r_{k}r_{k'}, {1}/{\beta}), & \text{otherwise}\end{cases} \notag\\
x_{mn}^{(t)} & \sim \mathrm{Po}\left(\sum_{k = 1}^{K}\sum_{k' = 1}^{K}\lambda_{kk'} \phi_{nk}^{(t)}\phi_{mk'}^{(t)}\right),\notag  \\
b_{mn}^{(t)} & = \mathbbm{1}(x_{mn}^{(t)} \geq 1). \notag
\end{align}

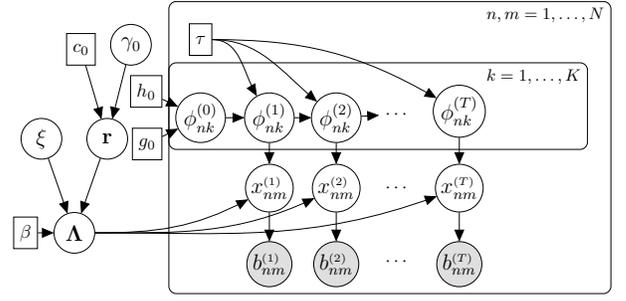
\begin{figure} 
  \centering
   \begin{tabular}{c} 
      \scalebox{.7}{
      \begin{tikzpicture}

  % Define nodes
  \node[latent]                  (bmn1) {$x_{nm}^{\scriptscriptstyle(1)}$};
  \node[latent, right=of bmn1, xshift=-0.65cm] (bmn2) {$x_{nm}^{\scriptscriptstyle(2)}$};
  \node[draw=none,right=of bmn2, xshift=-0.65cm]  (bmn3) {$\cdots$};
  \node[latent, right=of bmn3, xshift=-0.65cm] (bmnt)  {$x_{nm}^{\scriptscriptstyle(T)}$};
  % \node[draw=none,right=of bmnt, xshift=-0.65cm]  (bmnT) {$\cdots$};
  
  \node[obs, below= of bmn1, yshift=0.45cm]                  (amn1) {$b_{nm}^{\scriptscriptstyle(1)}$};
  \node[obs, below=of bmn2, , yshift=0.45cm] (amn2) {$b_{nm}^{\scriptscriptstyle(2)}$};
 \node[draw=none,below=of bmn3, yshift=0cm]  (amn3) {$\cdots$};
  \node[obs, below=of bmnt, yshift=0.45cm] (amnt)  {$b_{nm}^{\scriptscriptstyle(T)}$};
  % \node[draw=none,below=of bmnT, yshift=0cm]  (amnT) {$\cdots$};

  \node[latent, above=of bmn1, xshift=0cm, yshift=-0.6cm]  (phi_nm_1) {$\phi_{nk}^{(1)}$};
  \node[latent, above=of bmn2, xshift=0cm, yshift=-0.6cm]  (phi_nm_2) {$\phi_{nk}^{(2)}$};
   \node[draw=none,above=of bmn3]  (phi_nm_3) {$\cdots$};
  \node[latent, above=of bmnt, xshift=0cm, yshift=-0.5cm]  (phi_nm_t) {$\phi_{nk}^{(T)}$};
  % \node[draw=none,above=of bmnT]  (phi_nm_T) {$\cdots$};
   \node[latent, left=of phi_nm_1, xshift=0.65cm]  (phi_nm_0) {$\phi_{nk}^{(0)}$};
  \node[legend_isps, left=of phi_nm_0, yshift=-0.5cm,xshift=0.75cm](g0) {$g_0$};
  \node[legend_isps, left=of phi_nm_0, yshift= 0.5cm,xshift=0.75cm](h0) {$h_0$};
  
  \node[legend_isps, above=of phi_nm_0, yshift=-0.3cm,xshift= -0cm](cn) {$\tau$};
  
  \node[latent, left=of bmn1, xshift=-1.85cm,yshift=-0.85cm]  (lambda) {$\mathbf{\Lambda}$};
  \node[latent, above=of lambda, xshift=-0.625cm]  (xi) {$\xi$};
  \node[latent, above=of lambda, xshift=0.625cm]  (r_k) {$\mathbf{r}$};
  \node[latent, above=of r_k, xshift=0.45cm](gam0) {$\gamma_0$};
  \node[legend_isps, above=of r_k, xshift=-0.5cm](c0) {$c_0$};
%  \node[legend_isps, above=of xi, xshift=0.315cm](e0) {$e_0$};
%  \node[legend_isps, above=of xi, xshift=-0.315cm](f0) {$f_0$};
  \node[legend_isps, left=of lambda, xshift=0.7cm](beta) {$\beta$};
%  \node[latent, right=2cm of y]            (t) {$\tau$};

  % Connect the nodes
  \edge {bmn1} {amn1};
  \edge {bmn2} {amn2};

  \edge {bmnt} {amnt};

  \edge {g0, h0} {phi_nm_0};
  \edge {phi_nm_0} {phi_nm_1};
   \edge {phi_nm_1} {bmn1, phi_nm_2} ; %
   \edge {phi_nm_2} {bmn2, phi_nm_3} ; 
   \edge {phi_nm_t} {bmnt} ; 
   \edge {xi, r_k} {lambda} ; 
   \edge {gam0} {r_k};
   \edge {c0} {r_k};
%   \edge {e0} {xi};
%   \edge {f0} {xi};
   \edge {beta} {lambda};
   \draw[->,out=0, in=-152](lambda) to (bmn1);
   \draw[->,out=0, in=-155](lambda) to (bmn2);
   \draw[->,out=0, in=-160](lambda) to (bmnt);

   \draw[->,out=0, in=120](cn) to (phi_nm_1);
   \draw[->,out=0, in=140](cn) to (phi_nm_2);
   \draw[->,out=0, in=150](cn) to (phi_nm_t);

  % Plates
  \plate {yx} {(phi_nm_0)(phi_nm_1)(phi_nm_2)(phi_nm_3)(phi_nm_t)} {$k = 1,\ldots, K$} ;
  \plate {} {(cn)(phi_nm_0)(phi_nm_1)(phi_nm_2)(phi_nm_3)(phi_nm_t)(bmn1)(bmn2)(bmn3)(bmnt)(amn1)(amn2)(amn3)(amnt)} {$n, m= 1,\ldots, N$} ;

\end{tikzpicture}
      }
  \end{tabular}
  \caption{The graphical model of the Poisson gamma latent membership model; auxillary variables introduced for inference are not shown.}
  \label{gm}
\end{figure}

For our model's hyperparameters, we draw $c_0, \xi$ and $\beta$ from $\mathrm{Gam}(0.1, 1/0.1)$.
The graphical model is shown in Fig.~\ref{gm}. 

%% RELATED_WORK
\section{Related Work}

Approaches to analyze dynamic networks range from non-Bayesian methods such as the exponential random graph models~\cite{ERGM} or matrix and tensor factorization based methods~\cite{MTF} to Bayesian latent variable models~\cite{DIRM,DRIFT,EMMB,LFP,MGMM}. Our work falls into the latter class and we hence confine ourselves to discuss it's relation to this class.  
Dynamic extensions of \emph{mixed membership models}, where each node is assigned to a set of latent groups represented by multinomial distribution, have been developed~\cite{Fu09,EMMB}. One limitation of mixed membership models is that if the probability that node $i$ associates to group $k$ is increased, the probability that node $i$ associates to group $k'$ has to be decreased. The multi-group memberships models use a binary latent feature vector to characterize each node's multi-group memberships. In multi-group memberships models, a node's membership to one group does not limit its memberships to other groups. However, differences in the degree associations of a node to different groups cannot be captured by such models~\cite{DRIFT,LFP,MGMM}. 
One possible extension is to introduce a Gaussian distributed random variables to characterize how strongly each node is associated to different groups as previously done for latent factor models~\cite{Piyushi2008}. Such approaches where membership variables evolve according to linear dynamical systems~\cite{xing2010} can exploit the rich and efficient toolset for inference, such as Kalman filtering. However, the resulting signed-valued latent features lack an intuitive interpretation, e.g., in terms of degree of membership to a group. 
In contrast to these approaches, our model is based on a bilinear Poisson factor model~\cite{EPM}, where each node's memberships to groups are represented by a nonnegative real-valued memberships variable. The model does not only allow each node to be associated with multiple groups but also captures the degree at which each node is associated to a group. It means that our model combines the advantages of both mixed membership and multi-group memberships models. 
We exploit recent data augmentation technique~\cite{NBP}, to construct a sampling scheme for the time evolution of the nonnegative latent features. Related to our work is the dynamic gamma process Poisson factorization model (D-GPPF)~\cite{DGPPF} where the underlying groups' structure can evolve over time but each node-group membership is static. This is in constrast to our approach where the node's memberships evolve over time. We note that the gamma Markov chain used by our method and by D-GPPF is motivated by the augmentation techniques in ~\cite{NBP}.
In the experiment section we compare our model to (1) the hierarchical gamma process edge partition model (HGP-EPM)~\cite{EPM}, which is the static counterpart of our model,  (2) the dynamic relational infinite feature model (DRIFT)~\cite{DRIFT} which uses binary latent features to represent the node-group memberships, and characterizes the temporal dependences of latent features via a hidden Markov process and (3) the D-GPPF model.
%% INFERENCE
\section{Inference}
We present a Gibbs sampling procedure to draw samples of  $\{\phi_{nk}^{(t)}, \lambda_{kk'}, r_k, \xi, \gamma_0, \beta, c_0\}$ from their posterior distribution given the observed dynamic relational data and the hyper-parameters $\{\tau, g_0, h_0\}$. % \chk{inconsistent what you said earlier!!} 
In order to circumvent the technical challenges of drawing samples from the gamma Markov chain which does not yield closed-form posterior, we make use of the idea of data augmentation and marginalization technique and of the gamma-Poisson conjugacy to derive a closed-form update. 

\noindent\textbf{Notation.}
When expressing the full conditionals for Gibbs sampling we will use the shorthand ``--'' to denote all other variables or equivalently those in the Markov blanket for the respective variable according to Fig.~\ref{gm}.
We use ``$\cdot$'' as a index summation shorthand, e.g., $x_{\cdot j} = \sum_i x_{ij}$. 

We repeatedly exploit the following three results~\cite{PLBD,BLVM,NBP} to derive the conditional distributions used in our sampling algorithm.

\noindent\textbf{Result~1.}  
A negative binomially (NB) distributed random variable $y\sim\mathrm{NB}(r, p)$ can be generated from a gamma mixed Poisson distribution as, i.e., $y\sim \mathrm{Po}(\lambda)$ and $\lambda\sim \mathrm{Gam}(r, \frac{p}{1-p})$, as seen by marginalizing over $\lambda$. 

\noindent\textbf{Result~2.}
The Poisson-logarithmic bivariate distributed variable $(y,l)$ ~\cite{NBP} with $y\sim \mathrm{NB}(y;r, p)$ and a Chinese restaurant table (CRT)~\cite{csp} distributed variables  $l\sim\mathrm{CRT}(l;y, r)$, can equivalently be expressed as a sum-logarithmic (SumLog) and Poisson variable, i.e.,  
$y\sim\sum_{s=1}^{l}u_s$ with $u_s\sim\mathrm{Logarithmic}(p)$ and $l\sim\mathrm{Po}(-r\log(1-p))$.

\noindent\textbf{Result~3.}
Let $y_{\cdot} = \sum_{n=1}^{N}y_n$, where $y_n\sim \mathrm{Po}(\lambda_n)$ are independently drawn from a Poisson distribution with rate $\lambda_n$, then according to the Poisson-multinomial equivalence~\cite{BLVM}, we have $(y_1, \ldots, y_N)\sim \mathrm{Mult}(y_{\cdot};\frac{\lambda_1}{\sum_{n}\lambda_n}, \ldots, \frac{\lambda_N}{\sum_{n}\lambda_n})$ and $y_{\cdot}\sim\mathrm{Po}(\sum_{n}\lambda_n)$.

\subsection{Gibbs sampling}
\textbf{Sampling latent counts $x_{mn}^{(t)}$.}
We sample a latent count for each time dependent observed edge $b_{mn}^{(t)}$ as
\begin{align}\label{eq:lcount}
(x_{mn}^{(t)} \mid -) & \sim b_{mn}^{(t)}\mathrm{Po}_{+}\left(\sum_{k = 1}^{K}\sum_{k' = 1}^{K}\lambda_{kk'} \phi_{nk}^{(t)}\phi_{mk'}^{(t)}\right).
\end{align}

\noindent\textbf{Sampling individual counts $x_{mkk'n}^{(t)}$.}
% Using the Poisson additive property, 
We can partition the latent count \\$x_{mn}^{(t)} \sim \mathrm{Po}(\sum_{k,k' = 1}^{K}\lambda_{kk'} \phi_{nk}^{(t)}\phi_{mk'}^{(t)})$ using the Poisson additive property as\\ $x_{mn}^{(t)} = \sum_{k,k'}^{K}{x_{mkk'n}^{(t)}}$, where ${x_{mkk'n}^{(t)}}\sim\mathrm{Po}(\lambda_{kk'} \phi_{nk}^{(t)}\phi_{mk'}^{(t)})$. Then, via the  Poisson-multinomial equivalence, we sample the latent count $x_{mkk'n}^{(t)}$ as
\begin{align}\label{eq:la_int_count}
(x_{mkk'n}^{(t)}\mid -) & \sim \mathrm{Mult}\left(x_{mn}^{(t)} ; \frac{\lambda_{kk'} \phi_{nk}^{(t)}\phi_{mk'}^{(t)}}{\sum_{k = 1}^{K}\sum_{k' = 1}^{K}\lambda_{kk'} \phi_{nk}^{(t)}\phi_{mk'}^{(t)}}\right). 
\end{align}
\noindent\textbf{Sampling group weights $r_k$.}
Via the Poisson additive property, we have
\begin{align}\label{eq_x_k1k2}
x_{\cdot kk' \cdot}^{(\cdot)} & \sim \mathrm{Po}(\lambda_{kk'} \theta_{kk'}), 
\end{align}
where we defined $x_{\cdot kk'\cdot}^{(\cdot)} \equiv \sum_{t}\sum_{m,n\neq m} x_{mkk'n}^{(t)}$ and  $\theta_{kk'}\equiv\sum_{t}\sum_{n}\sum_{m \neq n}\phi_{nk}^{(t)}\phi_{mk'}^{(t)}$.
We can marginalize out $\lambda_{kk'}$ from Eq.~(\ref{eq_x_k1k2}) and (\ref{eq_pr_lambda}) using the gamma-Poisson conjugacy, which gives
\begin{align}
x_{\cdot kk' \cdot}^{(\cdot)} & \sim \mathrm{NB}(r_{k}\xi^{\delta_{kk'}}r_{k'}^{1-\delta_{kk'}}, \tilde{p}_{kk'}), \notag
\end{align}
where $\tilde{p}_{kk'} = \frac{\theta_{kk'}}{\theta_{kk'} + \beta}$ and $\delta_{kk'}$ denotes the Kronecker delta.
% $\delta_{kk'} = 1$  if $k=k'$, and $\delta_{kk'} =0$ otherwise. 
According to {Result}~2, we introduce the auxiliary variables as
\begin{align}\label{eq:l_kk}
l_{kk'}\sim \mathrm{CRT}(x_{\cdot kk' \cdot}^{(\cdot)}, r_{k}\xi^{\delta_{kk'}}r_{k'}^{1-\delta_{kk'}}).
\end{align}
We then re-express the bivariate distribution over $x_{\cdot kk' \cdot}^{(\cdot)}$ and $l_{kk'}$ as
\begin{align}% \label{eq_lkk}
x_{\cdot kk' \cdot}^{(\cdot)} & \sim \mathrm{SumLog}(l_{kk'}, r_{k}\xi^{\delta_{kk'}}r_{k'}^{1-\delta_{kk'}}), \notag\\
l_{kk'} & \sim \mathrm{Po}[-r_{k}\xi^{\delta_{kk'}}r_{k'}^{1-\delta_{kk'}}\log(1 - \tilde{p}_{kk'})]. \label{eq:l_kk_pois}
\end{align}
Using Eq.~(\ref{eq_pr_r}) and (\ref{eq:l_kk_pois}), via the gamma-Poisson conjugacy, we obtain the conditional distribution of $r_k$ as
\begin{align}
& (r_k \mid  - ) \sim \mathrm{Gam}\Big[\frac{\gamma_0}{K} + \sum_{k'}l_{kk'}, \notag \\
& \frac{1}{c_0 - \sum_{k'}\xi^{\delta_{kk'}}r_{k'}^{1-\delta_{kk'}}\log(1 - \tilde{p}_{kk'})}\Big]. \label{eq:r_k}
\end{align}
\noindent\textbf{Sampling intra-group weight $\xi$.}
We resample the auxiliary variables $l_{kk}$ using Eq.~(\ref{eq:l_kk}), and then exploit the gamma-Poisson conjugacy 
% with (\ref{eq_pr_xi}) and (\ref{eq:l_kk_pois}) 
to sample $\xi$ as
\begin{align}\label{eq:xi}
(\xi \mid  - ) &\sim \mathrm{Gam}\Big[0.1 + \sum_{k}l_{kk}, \frac{1}{0.1 - \sum_{k}r_k\log(1- \tilde{p}_{kk})}\Big]. 
\end{align}

\noindent\textbf{Sampling inter-group weights $\lambda_{kk'}$.}
We sample $\lambda_{kk'}$ from its conditional obtained from Eq.~(\ref{eq_pr_lambda}) and (\ref{eq_x_k1k2})  via the gamma-Poisson conjugacy as
%\begin{align}\label{eq:lambda_kk}
%(\lambda_{kk'} | - ) \sim \mathrm{Gam}\Big[x_{\cdot kk' \cdot}^{(\cdot)} + r_{k}& \xi^{\delta_{kk'}}r_{k'}^{1-\delta_{kk'}}, \notag \\ &{1}/{(\beta + \theta_{kk'}})\Big].
%\end{align}
\begin{align}\label{eq:lambda_kk}
(\lambda_{kk'} \mid  - ) \sim \mathrm{Gam}\Big[x_{\cdot kk' \cdot}^{(\cdot)} + r_{k}& \xi^{\delta_{kk'}}r_{k'}^{1-\delta_{kk'}},\frac{1}{(\beta + \theta_{kk'}})\Big].
\end{align}
\noindent\textbf{Sampling hyperparameter $\gamma_0$.}
Using Eq.~(\ref{eq:l_kk_pois}) and the Poisson additive property, we have $l_{k\cdot} \equiv \sum_{k'}l_{kk'}$ as
\begin{align}
l_{k\cdot} & \sim \mathrm{Po}[-r_{k}\sum_{k'}\xi^{\delta_{kk'}}r_{k'}^{1-\delta_{kk'}}\log(1 - \tilde{p}_{kk'})]. \notag
\end{align}
Marginalizing out $r_k$ using the gamma-Poisson conjugacy, we have
\begin{align}
l_{k\cdot} & \sim \mathrm{NB}({\gamma_0}/{K}, \hat{p}_{k}), \notag
\end{align}
where $\hat{p}_{k} = \frac{\sum_{k'}\xi^{\delta_{kk'}}r_{k'}^{1-\delta_{kk'}}\log(1 - \tilde{p}_{kk'})}{c_0 -\sum_{k'}\xi^{\delta_{kk'}}r_{k'}^{1-\delta_{kk'}}\log(1 - \tilde{p}_{kk'})}$. We introduce the auxiliary variables $\tilde{l}_k\sim\mathrm{CRT}(l_{k\cdot}, \gamma_0/K)$ and re-express the bivariate distribution over $l_{k\cdot}$ and $\tilde{l}_k$ as
\begin{align}
l_{k\cdot} & \sim \mathrm{SumLog}(\tilde{l}_k, \hat{p}_k), \notag\\
\tilde{l}_k&\sim\mathrm{Po}\Big[-\frac{\gamma_0}{K}\log(1- \hat{p}_k)\Big] \label{eq_lkt}.
\end{align}
%\begin{align}
%l_{k\cdot} & \sim \mathrm{SumLog}(\tilde{l}_k, \hat{p}_k), \qquad \tilde{l}_k\sim\mathrm{Po}\Big(-\frac{\gamma_0}{K}\log(1- \hat{p}_k)\Big) \label{eq_lkt}.
%\end{align}
Using Eq.~(\ref{eq_lkt}), we can then sample $\gamma_0$ via the gamma-Poisson conjugacy as
\begin{align}
(\gamma_0 \mid -) &\sim \mathrm{Gam}\Big[1 + \sum_k \tilde{l}_k, \frac{1}{1 - \frac{1}{K}\sum_k\log(1- \hat{p}_k) }\Big].\label{eq_gam0}
\end{align}

\noindent\textbf{Sampling latent memberships $\phi_{nk}^{(t)}$.}
% We adapt the data augmentation and marginalization technique developed in~\cite{NBP} and use the gamma-Poisson conjugacy to derive the closed-form update equations for $\phi_{nk}^{(t)}$. 
Since the latent memberships $\phi_{nk}^{(t)}$ evolve over time according to our Markovian construction, the backward and forward information need to be incorporated into the updates of $\phi_{nk}^{(t)}$.
We start from time slice $t=T$, 
%\begin{align}\label{}
%x_{nk\cdot\cdot}^{(T)} &\sim \mathrm{Po}(\phi_{nk}^{(T)}\omega_{nk}^{(T)}), \notag\\
%\phi_{nk}^{(T)} &\sim \mathrm{Gam}(\phi_{nk}^{(T-1)}/\tau, {1}/{\tau}),\notag
%\end{align}
\begin{align}\label{}
x_{nk\cdot\cdot}^{(T)} &\sim \mathrm{Po}(\phi_{nk}^{(T)}\omega_{nk}^{(T)}), \qquad \phi_{nk}^{(T)} \sim \mathrm{Gam}(\phi_{nk}^{(T-1)}/\tau, {1}/{\tau}),\notag
\end{align}
where
\begin{align}\label{}
x_{nk\cdot\cdot}^{(t)} & \equiv \sum_{\substack{m\neq n},k'} x_{mkk'n}^{(t)}, \qquad \omega_{nk}^{(t)} &\equiv \sum_{m\neq n, k'}\phi_{mk'}^{(t)}\lambda_{kk'}. \notag
\end{align}
Via the gamma-Poisson conjugacy, we have
\begin{align}\label{}
(\phi_{nk}^{(T)} \mid -) &\sim \mathrm{Gam}\Big[\phi_{nk}^{(T-1)}/\tau + x_{nk\cdot\cdot}^{(T)}, {1}/({\tau + \omega_{nk}^{(T)}})\Big].
\end{align}
%where
%\begin{align}\label{}
%x_{nk\cdot\cdot}^{(t)} & = \sum_{\substack{m\neq n},k'} x_{mkk'n}^{(t)}, \\
%\omega_{nk}^{(t)} &= \sum_{m\neq n, k'}\phi_{mk'}^{(t)}\lambda_{kk'}.
%% \varrho_{nk}^{(t)} &= \frac{\omega_{nk}^{(t)} - \log(1-\varrho_{nk}^{(t+1)})}{\tau + \omega_{nk}^{(t)} - \log(1-\varrho_{nk}^{(t+1)})}.
%\end{align}
Marginalizing out $\phi_{nk}^{(T)}$ yields
\begin{align}\label{}
x_{nk\cdot\cdot}^{(T)} &\sim \mathrm{NB}(\phi_{nk}^{(T-1)}/\tau, \varrho_{nk}^{(T)}),
\end{align}
where $\varrho_{nk}^{(T)} = \frac{\omega_{nk}^{(T)}}{\tau + \omega_{nk}^{(T)}}$.
According to Result~2, the NB distribution can be augmented with an auxiliary variable as
\begin{align}\label{}
y_{nk}^{(T)} &\sim \mathrm{CRT}(x_{nk\cdot\cdot}^{(T)}, \phi_{nk}^{(T-1)}/\tau).
\end{align}
We re-express the bivariate distribution over $x_{nk\cdot\cdot}^{(T)}$ and $y_{nk}^{(T)}$ as
\begin{align}
x_{nk\cdot\cdot}^{(T)} &\sim \mathrm{SumLog}(y_{nk}^{(T)}, \varrho_{nk}^{(T)}), \notag\\
y_{nk}^{(T)} &\sim \mathrm{Po}\Big[-\frac{\phi_{nk}^{(T-1)}}{\tau}\log(1 - \varrho_{nk}^{(T)})\Big]. \label{eq:y_nk_pois}
\end{align}
where 
\begin{align}
\varrho_{nk}^{(t)} = \frac{\omega_{nk}^{(t)} - \frac{1}{\tau}\log(1-\varrho_{nk}^{(t+1)})}{\tau + \omega_{nk}^{(t)} - \frac{1}{\tau}\log(1-\varrho_{nk}^{(t+1)})}.\label{eq:varrho}
\end{align}
Given $x_{nk\cdot\cdot}^{(T-1)}\sim\mathrm{Po}(\phi_{nk}^{(T-1)}\omega_{nk}^{(T-1)})$, via the Poisson additive property, we have
\begin{align}
y_{nk}^{(T)} + x_{nk\cdot\cdot}^{(T-1)} &\sim \mathrm{Po}\Big(\phi_{nk}^{(T-1)} \Big[ \omega_{nk}^{(T-1)}- \frac{1}{\tau}{\log(1 - \varrho_{nk}^{(T)})} \Big] \Big). \label{eq:backward}
\end{align}
Combing the likelihood in Eq.~(\ref{eq:backward}) with the gamma prior placed on $\phi_{nk}^{(T-1)}$, we immediately have the conditional distribution of $\phi_{nk}^{(T-1)}$ via the gamma-Poisson conjugacy as
\begin{align}\label{}
(\phi_{nk}^{(T-1)}|-) &\sim \mathrm{Gam}\Big[\phi_{nk}^{(T-2)}/\tau + y_{nk}^{(T)} + x_{nk\cdot\cdot}^{(T-1)}, \\ &\frac{1}{\tau + \omega_{nk}^{(T-1)} - \frac{1}{\tau}\log(1-\varrho_{nk}^{(T)})}\Big]. \notag
\end{align}
Here, $y_{nk}^{(T)}$ can be considered as the \emph{backward} information passed from $t = T$ to $T-1$. 
Recursively, we augment $\phi_{nk}^{(t)}$ at each time slice with an auxiliary variable $y_{nk}^{(t)}$  as
\begin{align}
y_{nk}^{(t+1)} + x_{nk\cdot\cdot}^{(t)} &\sim \mathrm{NB}(\phi_{nk}^{(t-1)}/\tau, \varrho_{nk}^{(t)}), \notag\\
y_{nk}^{(t)} &\sim \mathrm{CRT}(x_{nk\cdot\cdot}^{(t)} + y_{nk}^{(t+1)}, \phi_{nk}^{(t-1)}/\tau), \label{eq:y}
\end{align}
where the NB distribution over $y_{nk}^{(t+1)} + x_{nk\cdot\cdot}^{(t)}$ is obtained via the Poisson additive property and gamma-Poisson conjugacy with $x_{nk\cdot\cdot}^{(t)} \sim \mathrm{Po}(\phi_{nk}^{(t)}\omega_{nk}^{(t)})$. Repeatedly using Result~2, we have
\begin{align}
y_{nk}^{(t+1)} + x_{nk\cdot\cdot}^{(t)} &\sim \mathrm{SumLog}(y_{nk}^{(t)}, \varrho_{nk}^{(t)}), \notag\\
y_{nk}^{(t)} &\sim \mathrm{Po}\Big[-\frac{\phi_{nk}^{(t-1)}}{\tau}\log(1 - \varrho_{nk}^{(t)})\Big]. \notag
\end{align}
 By repeatedly exploiting the Poisson additive property and gamma-Poisson conjugacy, we obtain
\begin{align}
(\phi_{nk}^{(t-1)} \mid -) \sim \mathrm{Gam}&\Big[y_{nk}^{(t)} + \phi_{nk}^{(t-2)}/\tau + x_{nk\cdot\cdot}^{(t-1)}, \label{eq:phi}\\
& \frac{1}{\tau + \omega_{nk}^{(t-1)} - \frac{1}{\tau}\log(1-\varrho_{nk}^{(t)})}\Big]. \notag
\end{align}
%\begin{align}\label{}
%x_{nk\cdot\cdot}^{(t)} &\sim \mathrm{Po}(\phi_{nk}^{(t)}\omega_{nk}^{(t)}). \notag\\
%y_{nk}^{(t)} &\sim \mathrm{CRT}(x_{nk\cdot\cdot}^{(t)} + y_{nk}^{(t+1)}, \phi_{nk}^{(t-1)}). \notag
%\end{align}
%given the backward information from $t+1$ to $t$, $y_{nk}^{(t+1)}$, we have
%\begin{align}\label{}
%y_{nk}^{(t+1)} &\sim \mathrm{Po}(-\phi_{nk}^{(t)}\log(1 - \varrho_{nk}^{(t+1)})), \\
%x_{nk\cdot\cdot}^{(t)} &\sim \mathrm{Po}(\phi_{nk}^{(t)}\omega_{nk}^{(t)}).
%\end{align}
%Combing the gamma prior imposed on $\phi_{nk}^{(t)}$, we can marginalize over $\phi_{nk}^{(t)}$ as
%\begin{align}\label{}
%y_{nk}^{(t+1)} + x_{nk\cdot\cdot}^{(t)} &\sim \mathrm{NB}(\phi_{nk}^{(t-1)}, \varrho_{nk}^{(t)}),
%\end{align}
We sample the auxiliary variables $y_{nk}^{(t)}$ and update $\varrho_{nk}^{(t)}$ recursively from $t=T$ to $t=1$, which can be considered as the \emph{backward filtering} step. Then, in the \emph{forward} pass we sample $\phi_{nk}^{(t)}$ from $t = 1$ to $t=T$. 
\\
\noindent\textbf{Sampling hyperparameters.}
Via the gamma-gamma conjugacy, we sample $c_0$ and $\beta$ as
\begin{align}
(c_{\scriptscriptstyle 0}\mid-)&\sim\mathrm{Gam}\Big[0.1+\gamma_0, 1/(0.1+\sum_kr_k)\Big], \label{eq:hyper} \\
(\beta\mid-)&\sim\mathrm{Gam}\Big[0.1+\sum_{k,k'}r_k\xi^{\delta_{kk'}}r_{k'}^{1-\delta_{kk'}},\frac{1}{0.1+\sum_{\scriptscriptstyle k,k'}\lambda_{kk'}}\Big] \notag
\end{align} 
Algorithm~\ref{alg:batch-Gibbs} summarizes the full procedure.
%\begin{alignat}{2}
%y_{nk}^{(t)} \sim \mathrm{CRT}&\Big(y_{nk}^{(t+1)} + x_{nk\cdot\cdot}^{(t)}, \phi_{nk}^{(t-2)}/\tau\Big), \label{eq:y}\\
%\phi_{nk}^{(t)} \sim \mathrm{Gam}&\Big[y_{nk}^{(t+1)} + \phi_{nk}^{(t-1)}/\tau + x_{nk\cdot\cdot}^{(t)}, \label{eq:phi}\\
%& \frac{1}{\tau + \omega_{nk}^{(t)} - \frac{1}{\tau}\log(1-\varrho_{nk}^{(t+1)})}\Big]. \notag
%\end{alignat}

\begin{algorithm}[t]
    \DontPrintSemicolon
    \SetAlgoLined
    \SetKwInOut{KwInput}{input}
    \SetKwInOut{KwOutput}{output}
    \KwInput{dynamic relational data $\mathbf{b}^{(1)}, \ldots, \mathbf{b}^{(T)}$}
    Initialize the maximum number of groups $K$, hyperparameters $\beta, \tau, c_0, g_0, h_0$, and parameters $\gamma_0, r_k, \xi, \lambda_{kk'}, \phi_{nk}^{(t)}$\;
    \Repeat{convergence}{
    Sample $x_{mn}^{(t)}$ for non-zero links (Eq.~\ref{eq:lcount})\;
    Sample $x_{mkk'n}^{(t)}$ (Eq.~\ref{eq:la_int_count}) and update\;
    \quad $x_{\cdot kk'\cdot}^{(\cdot)} = \sum_{t}\sum_{m,n\neq m} x_{mkk'n}^{(t)}$\;
    \quad $x_{mk\cdot\cdot}^{(t)} = \sum_{\substack{n\neq m},k'} x_{mkk'n}^{(t)}$\;
    Sample $l_{kk'}$ (Eq.~\ref{eq:l_kk}) and calculate the quantities:\;
    \quad $\theta_{kk'}=\sum_{t, n, m \neq n}\phi_{nk}^{(t)}\phi_{mk'}^{(t)}$,\quad $\tilde{p}_{kk'} = \frac{\theta_{kk'}}{\theta_{kk'} + \beta}$\;
    Sample $r_k$ (Eq.~\ref{eq:r_k}), $\xi$ (Eq.~\ref{eq:xi}), and $\lambda_{kk'}$ (Eq.~\ref{eq:lambda_kk})\;
     %\tcc{backward filtering}
    \For{$t = T$ \emph{to} $1$}{
%    Update $\hat{p}_{k} = \frac{\sum_{k'}\xi^{\delta_{(k,k')}}(r_{k'})^{1-\delta_{(k,k')}}\log(1 - \tilde{p}_{kk'})}{c_0 -\sum_{k'}\xi^{\delta_{(k,k')}}(r_{k'})^{1-\delta_{(k,k')}}\log(1 - \tilde{p}_{kk'})}$\;
    Sample $y_{nk}^{(t)}$ (Eq.~\ref{eq:y}) and update $\varrho_{nk}^{(t)}$ (Eq.~\ref{eq:varrho})\;
    }
    %\tcc{forward sampling}
     \For{$t = 0$ \emph{to} $T$}{
    Sample $\phi_{nk}^{(t)}$ (Eq.~\ref{eq:phi})\;
    }
    Sample $c_0$, $\beta$ (Eq.~\ref{eq:hyper}) and $\gamma_0$~(Eq.~\ref{eq_gam0})\;
    % Sample hyperparameters $\beta$, $\tau, c_0, e_0, f_0, g_0, h_0$\;
%      Draw a single sample $\bz\sim q(\bz;\bv)$\;
%      Compute the auxiliary functions $\hfun\left(\Tcal^{-1}(\bz;\bv);\bv\right)$ and $\ufun\left(\Tcal^{-1}(\bz;\bv);\bv\right)$ (Eq.~\ref{eq:hfun_ufun})\;
%      Estimate $\bg^{\textrm{rep}}$ and $\bg^{\textrm{corr}}$ (Eq.~\ref{eq:grep_gcorr}, estimate the expectation with one sample)\;
%      Compute (analytic) or estimate (Monte Carlo) the gradient of the entropy, $\nabla_{\bv}\ent{q(\bz;\bv)}$\;
%      Compute the noisy gradient $\nabla_{\bv}\Lcal$ (Eq.~\ref{eq:corrected_reparam})\;
%      Set the step-size $\brho^{(i)}$ (Eq.~\ref{eq:step_schedule}) and take a gradient step for $\bv$\; % (Eq.~\ref{eq:v_new})\;
    }
    \KwOutput{posterior mean $\phi_{nk}^{(t)}, r_k, \xi, \lambda_{kk'}, \beta, \gamma_0,c_0$}
    \caption{Batch Gibbs Sampling\label{alg:batch-Gibbs}}
\end{algorithm}
\subsection{Online Gibbs sampling}
To make our model applicable to large-scale dynamic networks, we propose an online Gibbs sampling algorithm based on the recent developed Bayesian conditional density filtering (BCDF)~\cite{BCDF}, which has been adapted for 
% knowledge graph learning~\cite{TopicKG} and 
Poisson tensor factorization~\cite{PTF} recently. The main idea of BCDF is to partition the data into small mini-batches, and then to perform inference by updating the sufficient statistics using each mini-batch in each iteration. Specifically, the sufficient statistics used in our model are the latent count numbers. We use $J$ and $J^{i}$ to denote the indices of the entire data and the mini-batch in $i\mbox{-}\mathrm{th}$ iteration respectively. We define the quantities updated with the mini-batch in $i\mbox{-}\mathrm{th}$ iteration as:
%\begin{align}\label{}
%x_{mk\cdot\cdot}^{(t) i} & = \frac{|J|}{|J^{i}|}\sum_{\substack{n\neq m,\\ m, n \in {\scriptscriptstyle J_t}}}\sum_{k'} x_{mkk'n}^{(t)},\notag \\
%x_{\cdot kk'\cdot}^{i} & = \frac{|J|}{|J^{i}|}\sum_t\sum_{\substack{m,n\neq m,\\ m, n \in {\scriptscriptstyle J_t}}} x_{mkk'n}^{(t)}.\notag
%\end{align}
\begin{align}\label{}
x_{mk\cdot\cdot}^{(t) i} = \frac{|J|}{|J^{i}|}\sum_{\substack{k',n\neq m,\\ m, n \in J^i}} x_{mkk'n}^{(t)}, \ \ 
x_{\cdot kk'\cdot}^{i} = \frac{|J|}{|J^{i}|}\sum_{\substack{t,m,n\neq m,\\ m, n \in J^i}} x_{mkk'n}^{(t)}.\notag
\end{align}
The main procedure of our online Gibbs sampler is then as follows.
We first update the sufficient statistics used to sample model parameters as
\begin{align}\label{}
x_{mk\cdot\cdot}^{(t)i} & = (1-\rho^{i})x_{mk\cdot\cdot}^{i-1} + \rho^{i}\frac{|J|}{|J^{i}|}\sum_{\substack{n\neq m,\\ m, n \in {\scriptscriptstyle J^i}}}\sum_{k'} x_{mkk'n}^{(t)},\notag \\
x_{\cdot kk'\cdot}^{i} & = (1-\rho^{i})x_{\cdot kk'\cdot}^{i} + \rho^{i}\frac{|J|}{|J^{i}|}\sum_{t}\sum_{\substack{m,n\neq m,\\ m, n \in {\scriptscriptstyle J^i}}} x_{mkk'n}^{(t)},\notag
\end{align}
where $\rho^{i} = (i + i_{0})^{-\kappa}$, where $i_0>0$ and $\kappa \in (1/2, 1]$ is the decay factor commonly used for online methods~\cite{BCDF}. 
We calculate the sufficient statistics for each mini-batch and then resample the model parameters using the procedure in batch Gibbs sampling algorithm outlined in Algorithm~\ref{alg:batch-Gibbs}.

%% EXPERIMENTS
\section{Experiments}
We evaluate our model by performing experiments on both synthetic and real-world datasets. First, we generate a synthetic data with the true underlying network structure evolving over time to test our model on dynamic community detection.
%, follwing\cite{DRIFT, LFP, MGMM}. 
For quantitive evaluation, we determine the model's ability to predict held-out missing links. Our baseline methods include DRIFT, D-GPPF and HGP-EPM as we discussed before.
% We compare the model with two baseline methods: (1) the dynamic relational infinite feature model (DRIFT)~\cite{DRIFT} which uses binary latent features to represent the node-group memberships, and characterizes the temporal dependences of latent features via a hidden Markov process; (2) the dynamic gamma process Poisson factorization (D-GPPF) model~\cite{DGPPF}, which is similar in construction to our model but cannot explicitly capture the changes of the latent node memberships over time. 
For DRIFT, we use default settings as the code released online.~\footnote{http://jfoulds.informationsystems.umbc.edu/code/DRIFT.tar.gz.} We implemented D-GPPF by ourselves and set the hyperparameters and initialize the model parameters with the values provided in~\cite{DGPPF}. For HGP-EPM, we used the code released for~\cite{EPM}.~\footnote{https://github.com/mingyuanzhou/EPM.} In the following, we refer to our model as DPGM (\textbf{D}ynamic \textbf{P}oisson \textbf{G}amma \textbf{M}embership model). For DPGM, we set $\tau = 1, g_0=0.1, h_0=0.1$ and use $K=N/2$, where $N$ is the number of nodes, for initilization. 
We obtain similar results when instead setting $\tau=0.1, \tau=10$ in a sensitivity analysis.
For online Gibbs sampling, we set $\kappa = 0.5, i_0 = 100$, and mini-batch size $|J_i| = N/4$. 
All our experiments were performed on a standard desktop with 2.7 GHz CPU and 24 GB RAM.
% The source code of our model (DPGM) will be released upon publication.
Following~\cite{DRIFT}, we generate a set of small-scale dynamic networks from real-world relational data while we use held-out relational data to evaluate our model. The following three real-world datasets are used in our experiments, the detail of which are summarized in Table~\ref{datasets}.
%\noindent

\textbf{NIPS.} The dataset records the co-authorship information among 5722 authors on publications in NIPS conferences over the past ten years~\cite{nips_coauthor}. We first take the 70 authors who are most connected across all years to evaluate all methods (NIPS 70). We also use the whole dataset (NIPS 5K) for evaluation.

\textbf{Enron.} The dataset contains 136776 emails among 2359 persons over 28 months ($T = 28$)~\cite{Enron}. We generate a binary symmetric matrix for each monthly snapshot. The presence or absence of an email between each pair of persons during one month is described by the binary link at that time. We first select 61 persons by taking a 7-core of the aggregated network for the entire time and filter out the authors with email records less than 5 snapshots (Enron 61). We also use the whole dataset for evaluation (Enron 2K). 

\textbf{DBLP.} The DBLP dynamic networks dataset~\cite{Enron} are generated from the co-authorship recordings among 347013 authors over 25 years, which is a subset of data contained in the DBLP database. We first choose 7750 authors by taking a 7-core of the aggregated network for the entire time (DBLP 7K) and subsequently filter out authors with less than 10 years of consecutive publication activity to generate a small dataset (DBLP 96).
The proposed method and the two baselines are applied to all six datasets, except for DRIFT. DRIFT could not be applied to NIPS 5K, Enron 2K and DBLP 7K due to its unfavorable computational complexity. Most of these datasets exhibit strong sparsity that the proposed algorithm can exploit through its Bernoulli-Poisson link function.   
\begin{table}[h]
\small
\centering
\caption{\label{datasets} {Details of the dataset used in our experiments.}}
% \vspace{-1.0em}
\begin{tabular}{|l|c|c|c|}\hline
Datasets & {NIPS 70} & {DBLP 96} & {Enron 61}\\ \hline %\cmidrule(r){2-7}%\cmidrule(l){8-15}
Nodes $\#$ &  70 & 96 & 61 \\\hline
Time slices $\#$ & 10 & 25 & 28  \\\hline
 %\vspace{0.2em}
Non-zero links $\#$ & 528 & 1392 & 1386 \\\hline
 \hline
 & {NIPS 5K} & {DBLP 7K} & {Enron 2K}\\ \hline %\cmidrule(r){2-7}%\cmidrule(l){8-15}
Nodes $\#$ &  5722 & 7750 & 2359 \\\hline
Time slices $\#$ & 10 & 10 & 28  \\\hline
Non-zero links $\#$ & 5514 & 108980 & 76828 \\\hline
\end{tabular}
\end{table}
\subsection{Dynamic community detection}

\begin{figure*} [htb]
%\begin{table}
  \centering
  \begin{tabular}{  m{0cm}  m{2.25cm}  m{2.25cm}   m{2.25cm}  m{2.25cm}  m{2.25cm}  m{2.25cm} m{0cm} } % { | m{-1cm}  c | c | c | c  m{-2cm} |}
   % \hline
  \vspace{-20em}
%  &  \scriptsize \ \ \ \ \ \ \ \ \ \ \ \ \ \ \ \ \ \ \ \ N = 20 & \scriptsize \ \ \ \ \ \ \ \ \ \ \ \ \ \ \ \ \ \ \ \ N = 40 & \scriptsize \ \ \ \ \ \ \ \ \ \ \ \ \ \ \ \ \ \ \ \ N = 60 & \scriptsize \ \ \ \ \ \ \ \ \ \ \ \ \ \ \ \ \ \ \ \ N = 100 &\\ %\hline \vspace{-0.5em} 
    %\multirow{3}{*}{\rotatebox[origin=c]{90}{Text}} & 
     % \rotatebox[origin=Bc]{90}{\parbox{2.8cm}{{\scriptsize \ \ \ \ \ \ \ \ \ }}}
    &
      \begin{minipage}{.21\textwidth}
      \includegraphics[width=2.8cm,height=1.5cm, keepaspectratio]{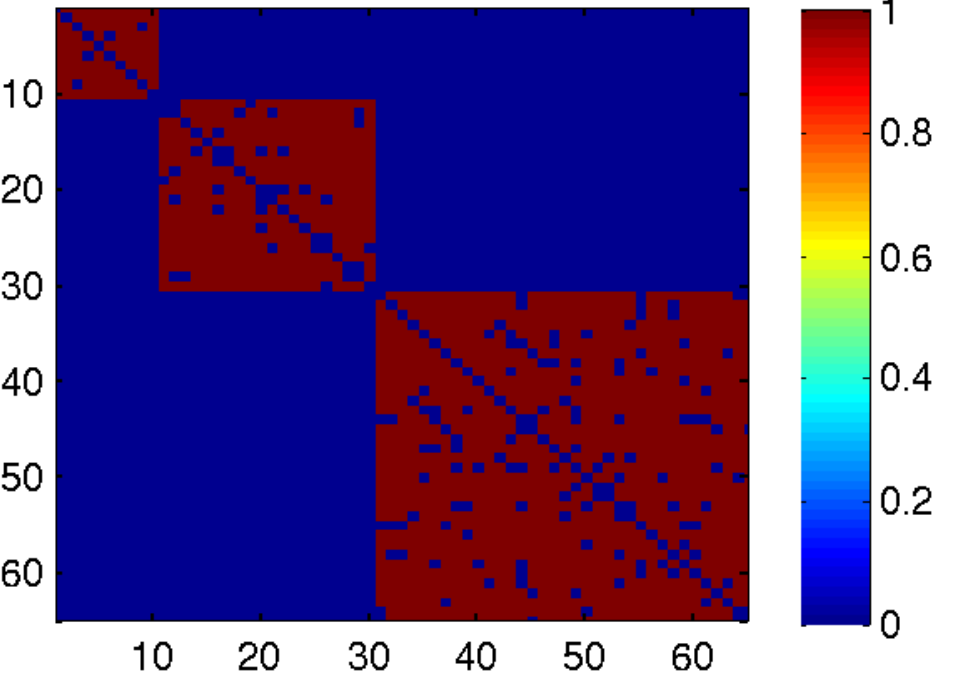}
    \end{minipage}
    &
    %\begin{minipage}[t]{5cm}
      \begin{minipage}{.21\textwidth}
      \includegraphics[width=2.8cm,height=1.5cm, keepaspectratio]{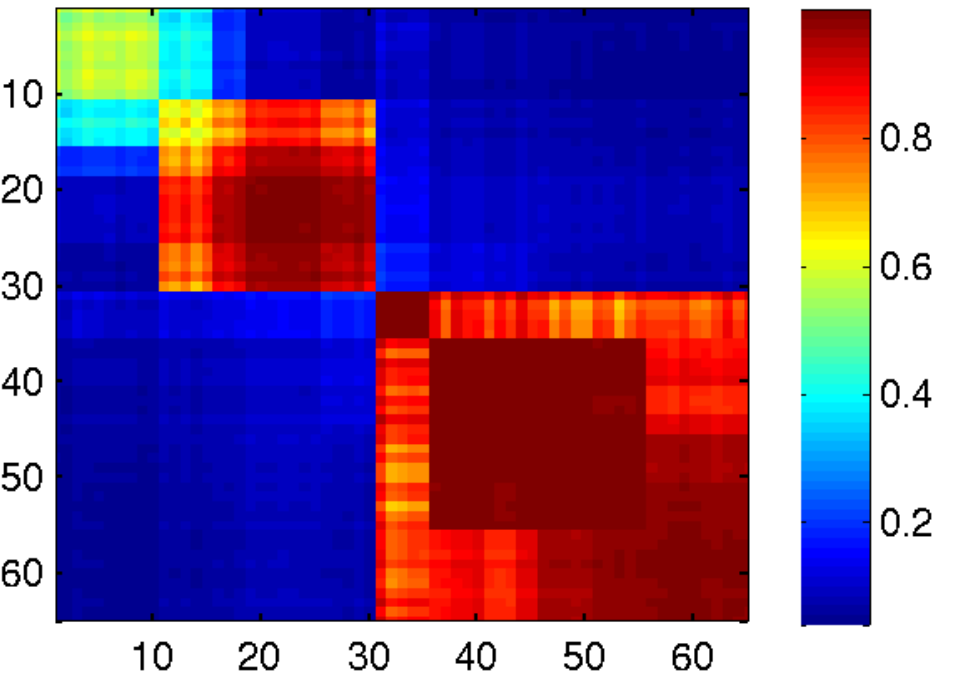}
    \end{minipage}
    %\end{minipage}
    & 
    %\begin{minipage}{5cm}
      \begin{minipage}{.21\textwidth}
      \includegraphics[width=2.8cm,height=1.5cm, keepaspectratio]{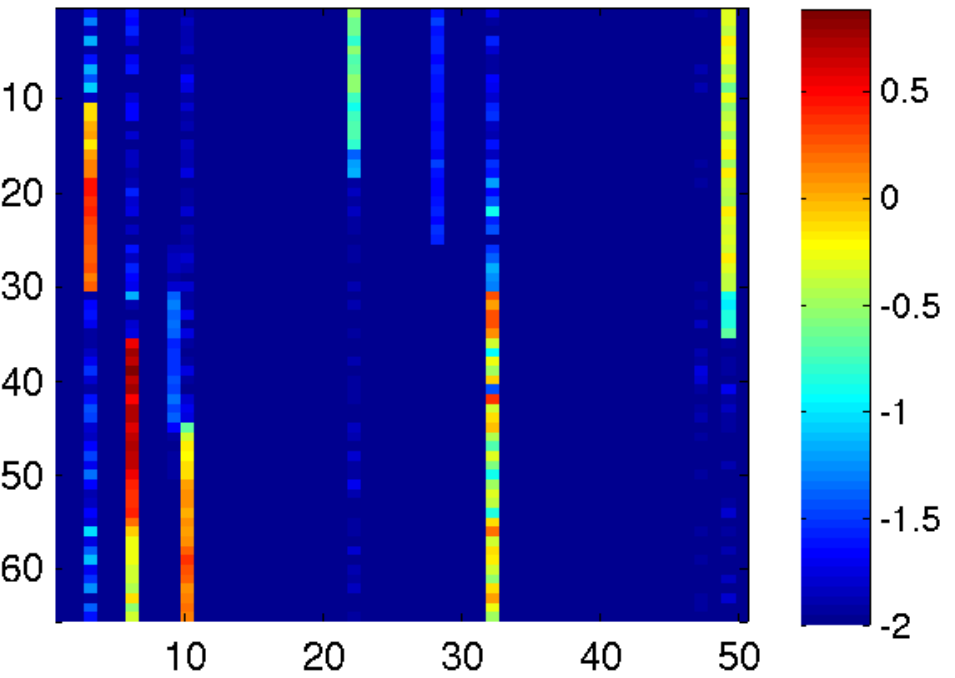}
    \end{minipage}
    %\end{minipage}
    &  
    %\begin{minipage}{5cm}
      \begin{minipage}{.21\textwidth}
      \includegraphics[width=2.8cm,height=1.5cm, keepaspectratio]{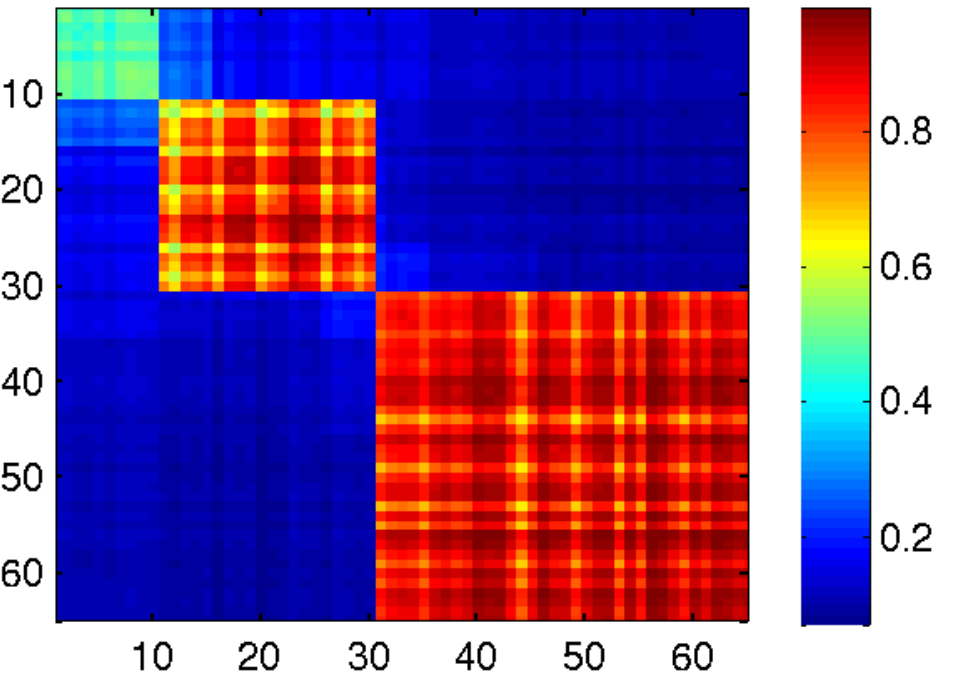}
    \end{minipage}
    &  
    %\begin{minipage}{5cm}
      \begin{minipage}{.21\textwidth}
      \includegraphics[width=2.8cm,height=1.5cm, keepaspectratio]{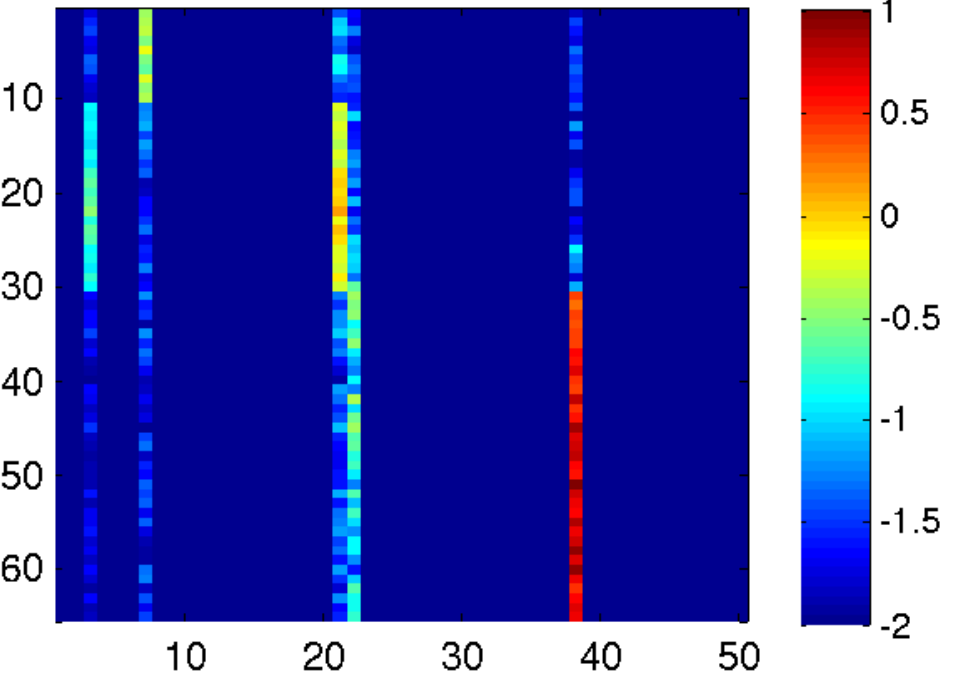}
    \end{minipage}
%    &  
%    %\begin{minipage}{5cm}
%      \begin{minipage}{.21\textwidth}
%      \includegraphics[width=2.8cm,height=1.5cm, keepaspectratio]{truth_t_6}
%    \end{minipage}
  \\
  \vspace{-20em}
  %\rotatebox[origin=Bc]{90}{\parbox{2.8cm}{{\scriptsize \ \ \ \ \ \ \ \ \  }}}
    &
      \begin{minipage}{.2\textwidth}
      \includegraphics[width=2.8cm,height=1.5cm, keepaspectratio]{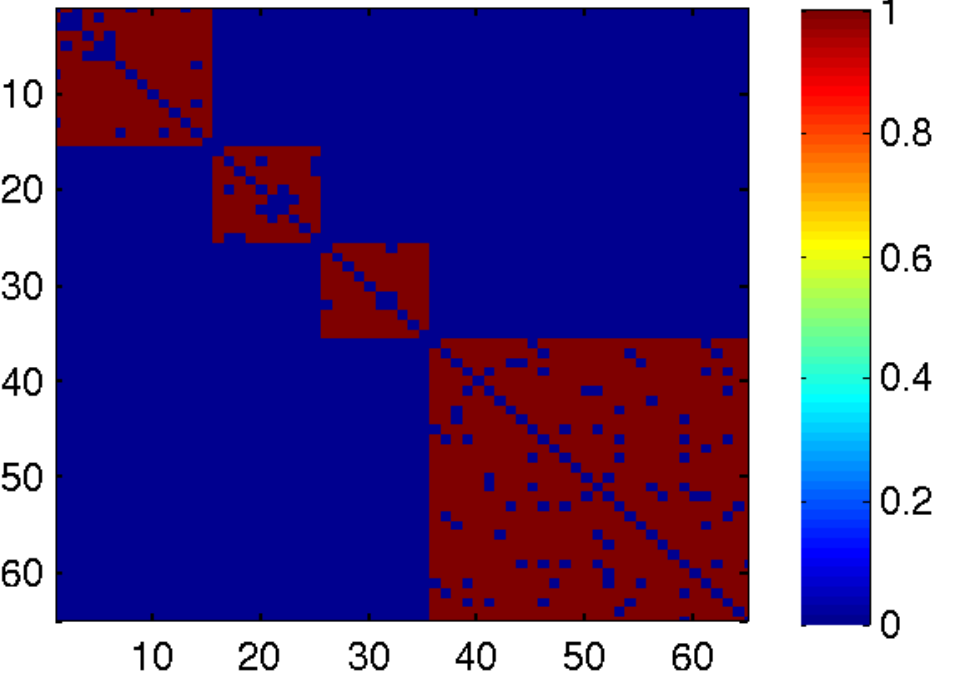}
    \end{minipage}
    &
    %\begin{minipage}[t]{5cm}
      \begin{minipage}{.21\textwidth}
      \includegraphics[width=2.8cm,height=1.5cm, keepaspectratio]{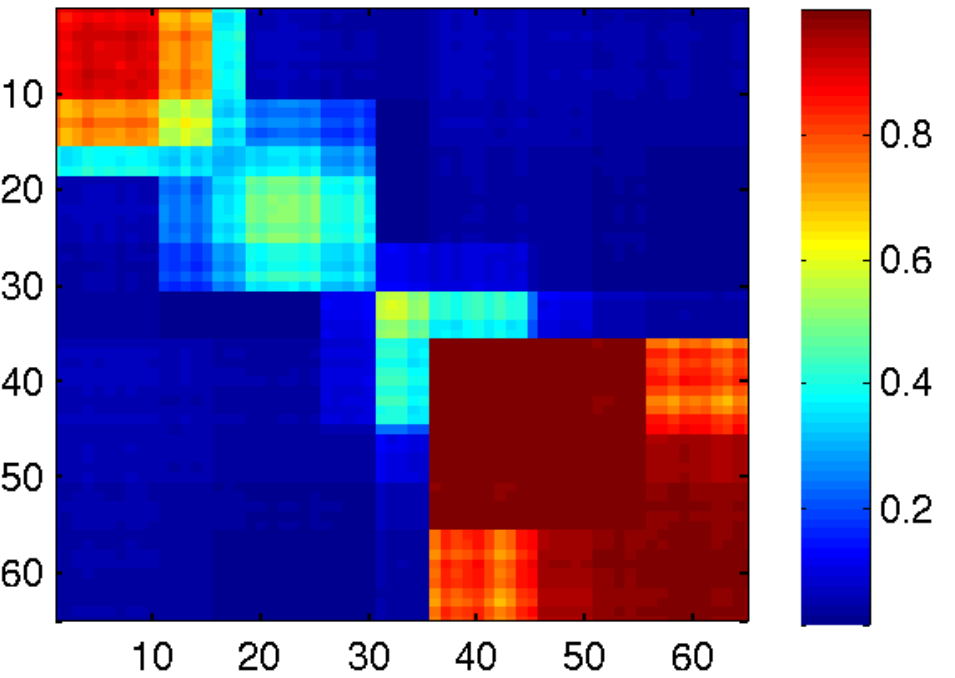}
    \end{minipage}
    %\end{minipage}
    & 
    %\begin{minipage}{5cm}
      \begin{minipage}{.21\textwidth}
      \includegraphics[width=2.8cm,height=1.5cm, keepaspectratio]{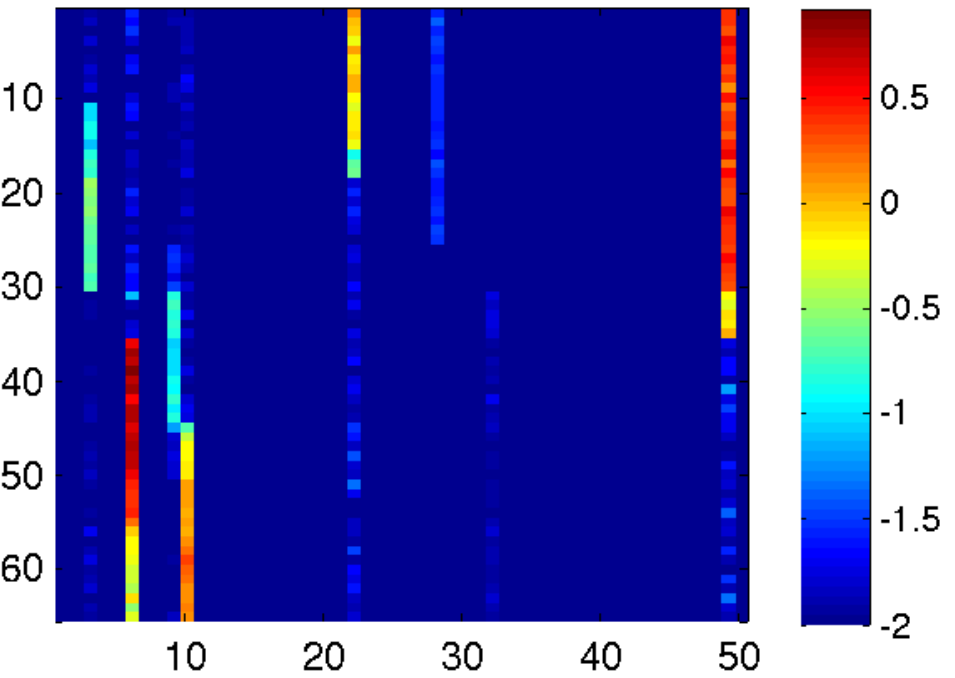}
    \end{minipage}
    %\end{minipage}
    &  
    %\begin{minipage}{5cm}
      \begin{minipage}{.21\textwidth}
      \includegraphics[width=2.8cm,height=1.5cm, keepaspectratio]{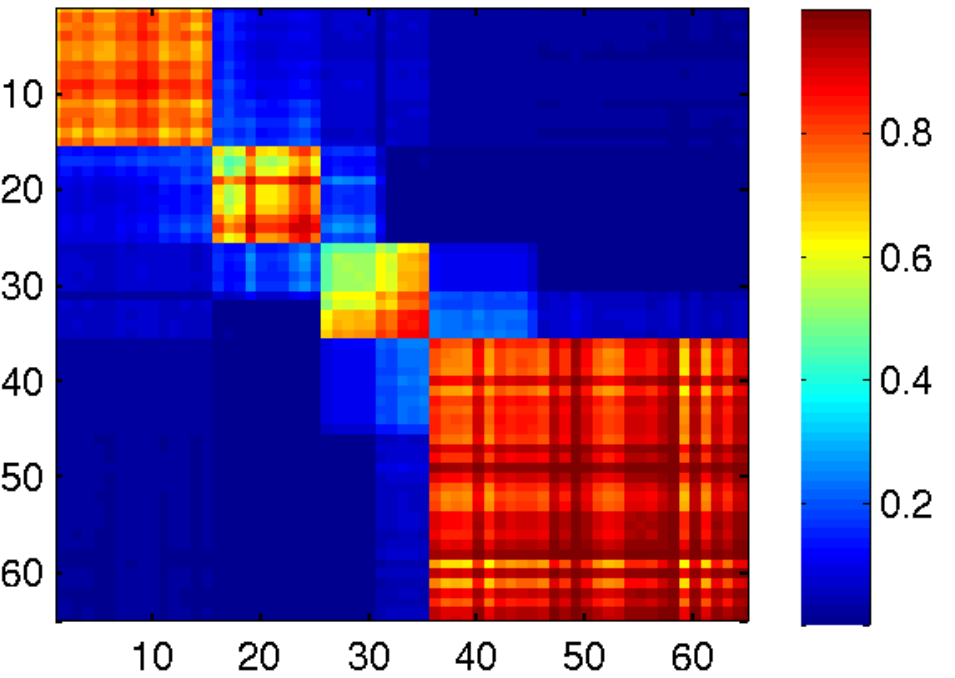}
    \end{minipage}
    &  
    %\begin{minipage}{5cm}
      \begin{minipage}{.21\textwidth}
      \includegraphics[width=2.8cm,height=1.5cm, keepaspectratio]{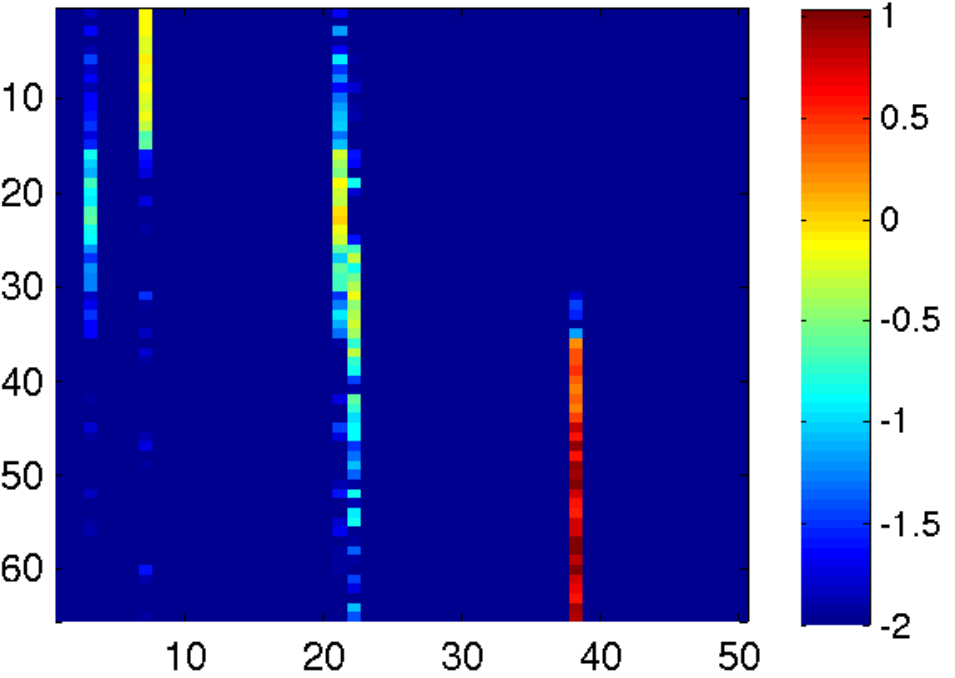}
    \end{minipage}
    \\
    \vspace{-20em}
    %\rotatebox[origin=Bc]{90}{\parbox{2.8cm}{{\scriptsize \ \ \ \ \ \ \ \ \  }}}
    &
      \begin{minipage}{.2\textwidth}
      \includegraphics[width=2.8cm,height=1.5cm, keepaspectratio]{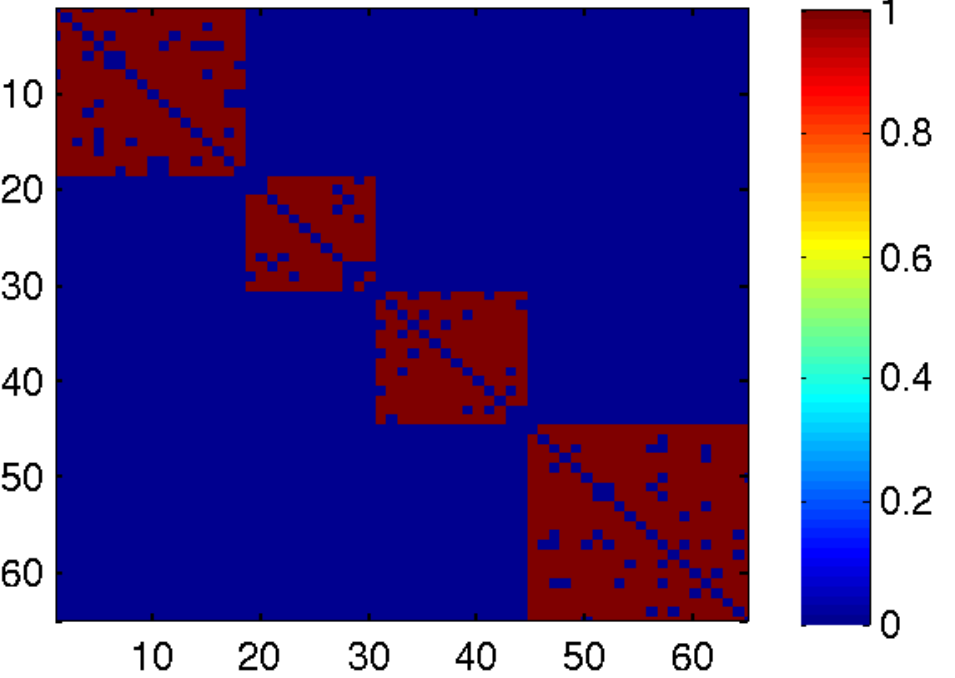}
    \end{minipage}
    &
    %\begin{minipage}[t]{5cm}
      \begin{minipage}{.21\textwidth}
      \includegraphics[width=2.8cm,height=1.5cm, keepaspectratio]{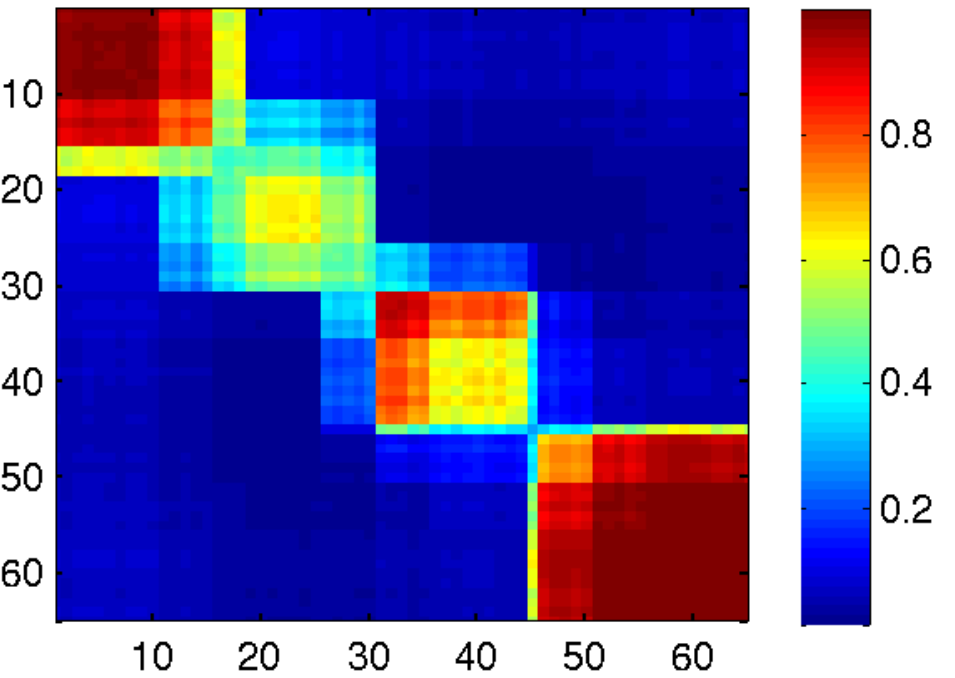}
    \end{minipage}
    %\end{minipage}
    & 
    %\begin{minipage}{5cm}
      \begin{minipage}{.21\textwidth}
      \includegraphics[width=2.8cm,height=1.5cm, keepaspectratio]{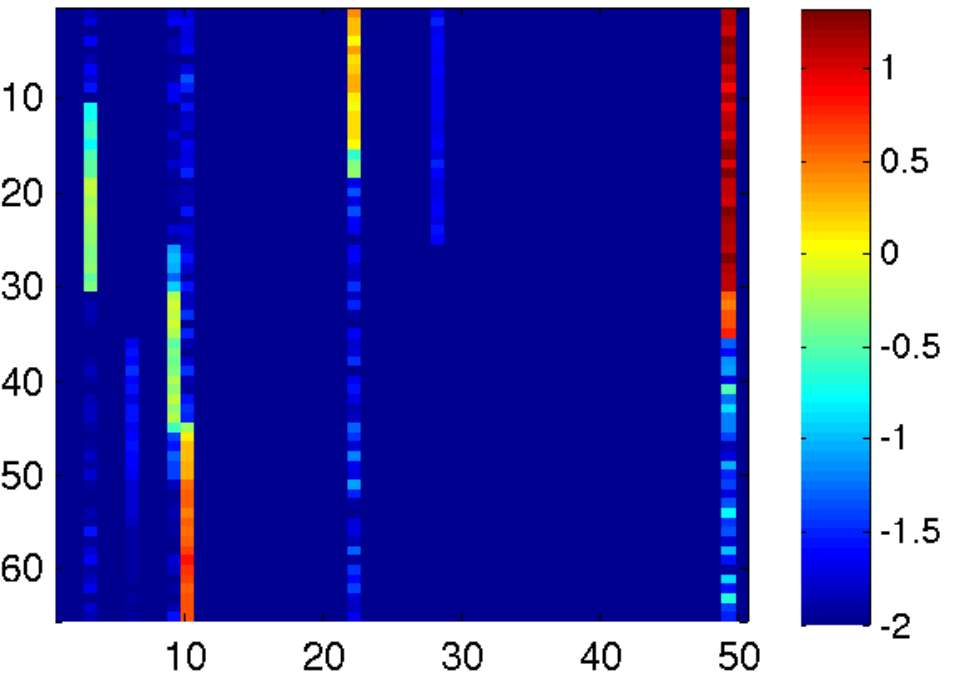}
    \end{minipage}
    %\end{minipage}
    &  
    %\begin{minipage}{5cm}
      \begin{minipage}{.21\textwidth}
      \includegraphics[width=2.8cm,height=1.5cm, keepaspectratio]{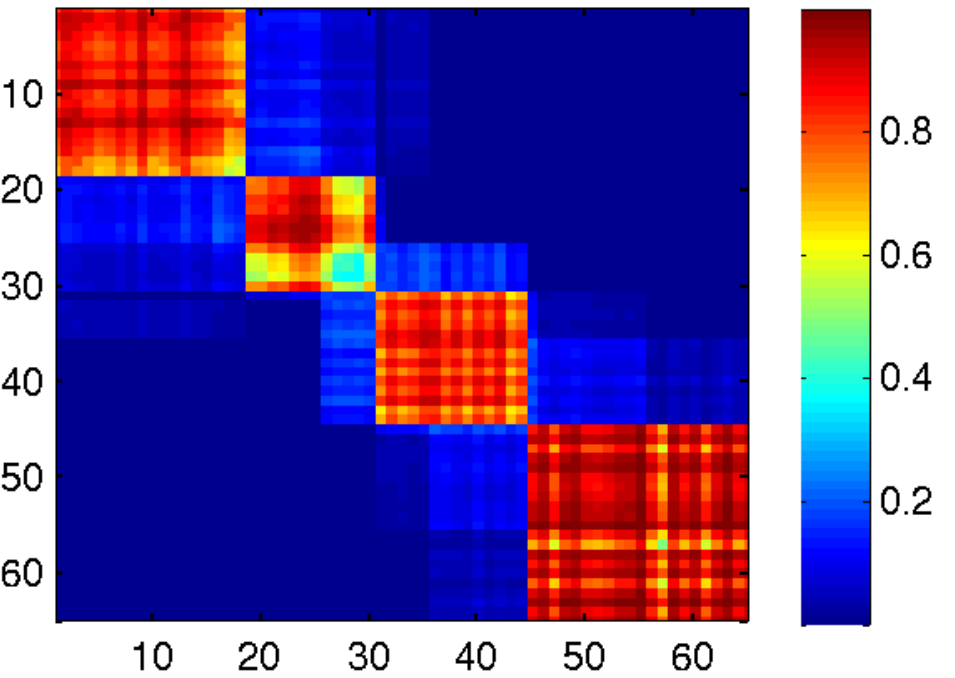}
    \end{minipage}
    &  
    %\begin{minipage}{5cm}
      \begin{minipage}{.21\textwidth}
      \includegraphics[width=2.8cm,height=1.5cm, keepaspectratio]{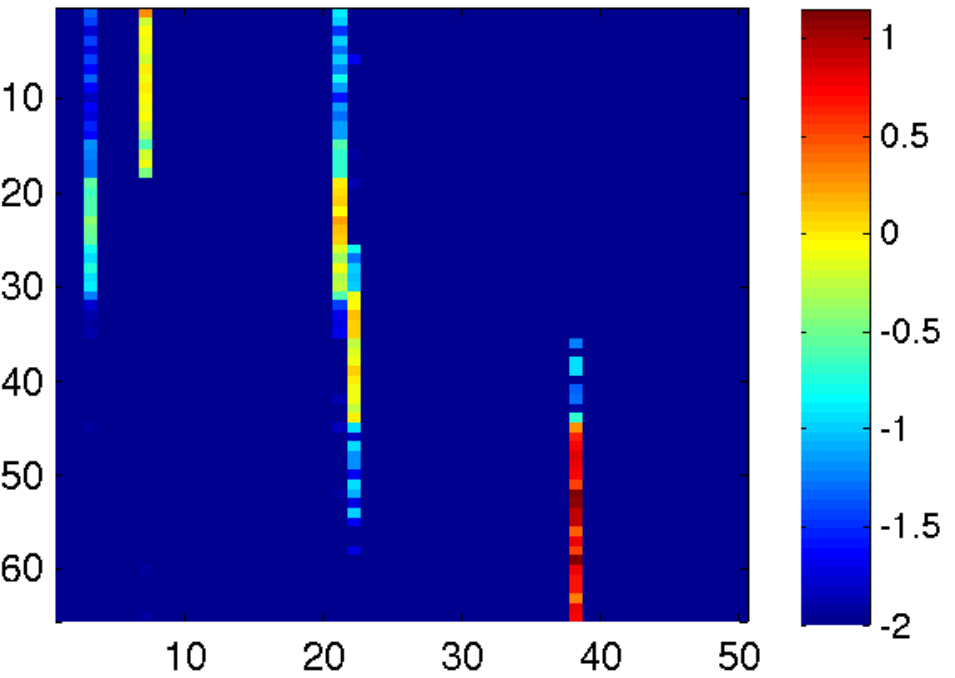}
    \end{minipage}
    \\
    \vspace{-2em}
   % \rotatebox[origin=Bc]{90}{\parbox{2.8cm}{{\scriptsize \ \ \ \ \ \ \ \ \  }}}
    &
      \begin{minipage}{.2\textwidth}
      \includegraphics[width=2.8cm,height=1.5cm, keepaspectratio]{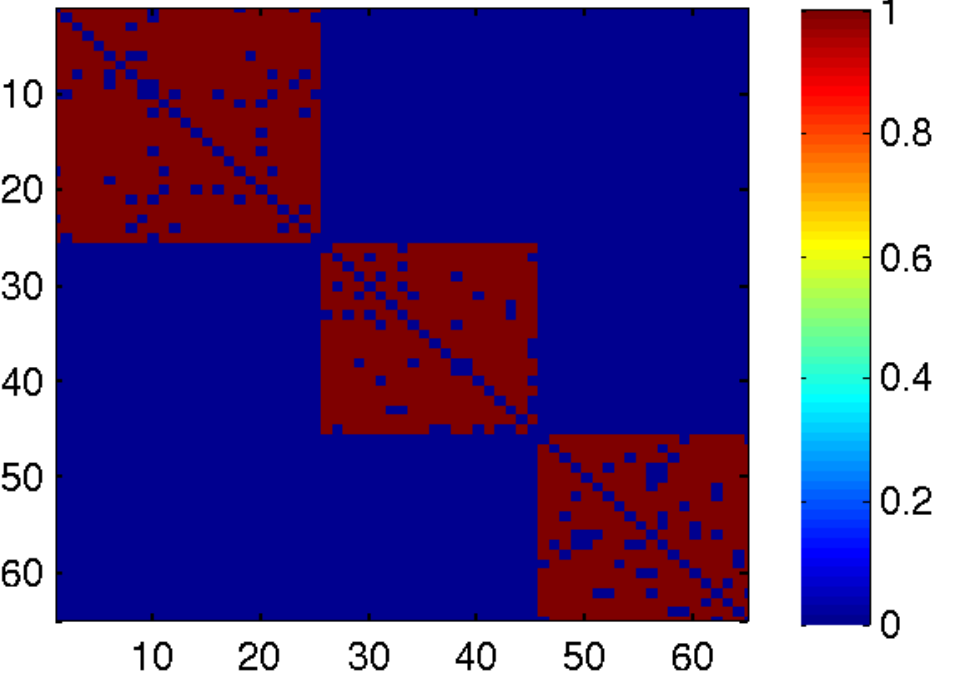}
    \end{minipage}
    &
    %\begin{minipage}[t]{5cm}
      \begin{minipage}{.21\textwidth}
      \includegraphics[width=2.8cm,height=1.5cm, keepaspectratio]{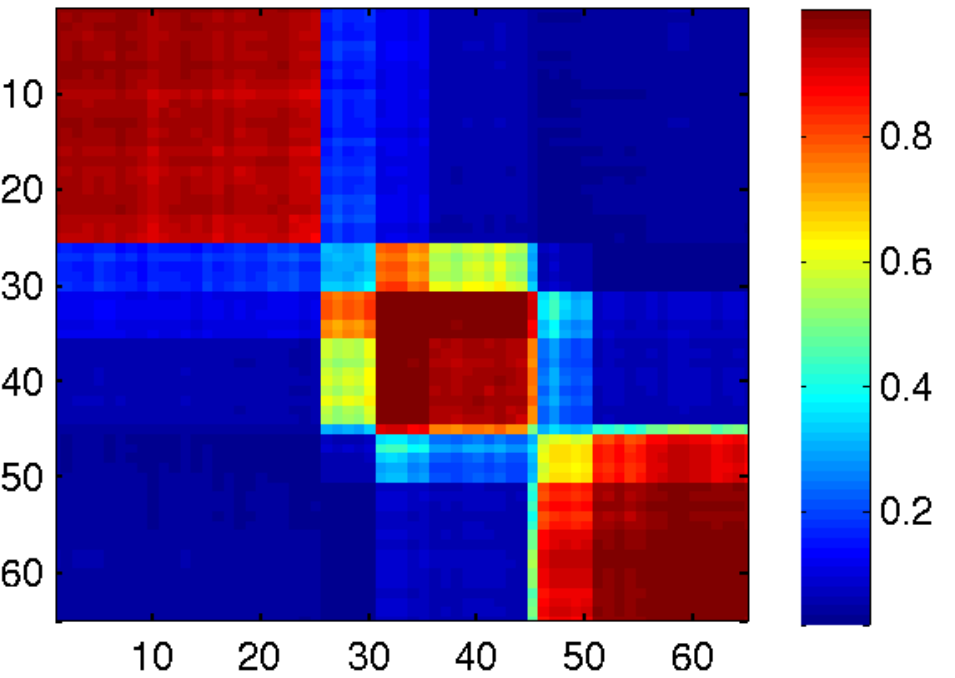}
    \end{minipage}
    %\end{minipage}
    & 
    %\begin{minipage}{5cm}
      \begin{minipage}{.21\textwidth}
      \includegraphics[width=2.8cm,height=1.5cm, keepaspectratio]{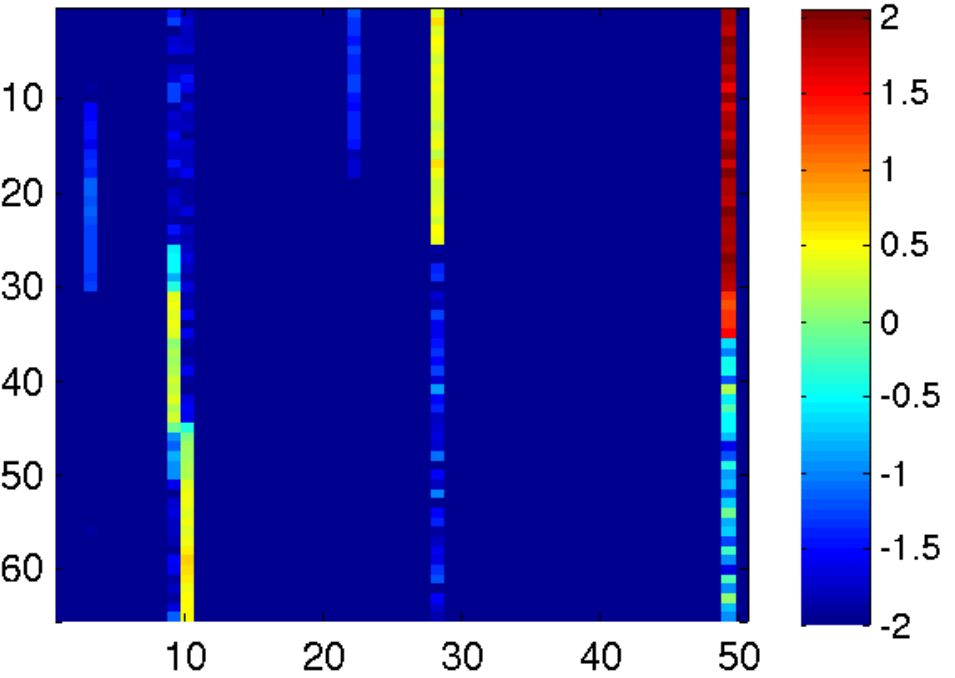}
    \end{minipage}
    %\end{minipage}
    &  
    %\begin{minipage}{5cm}
      \begin{minipage}{.21\textwidth}
      \includegraphics[width=2.8cm,height=1.5cm, keepaspectratio]{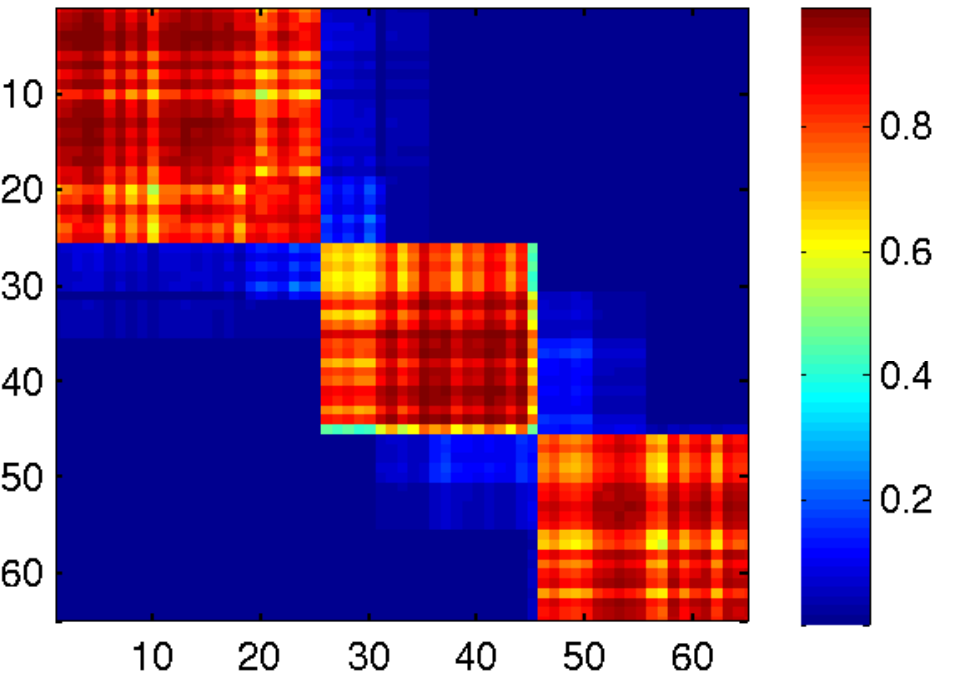}
    \end{minipage}
    &  
    %\begin{minipage}{5cm}
      \begin{minipage}{.21\textwidth}
      \includegraphics[width=2.8cm,height=1.5cm, keepaspectratio]{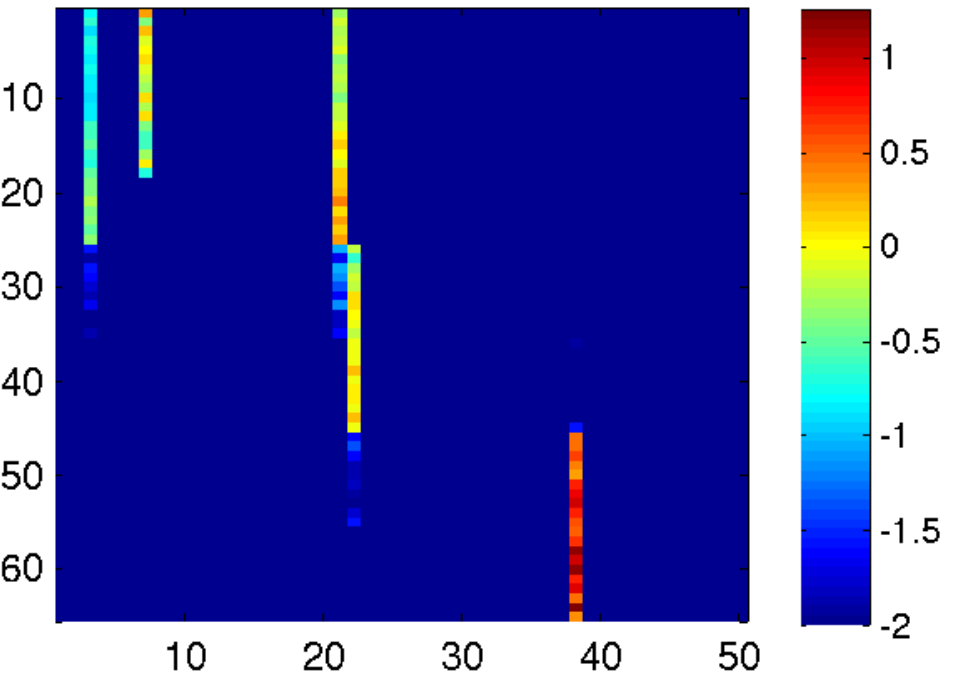}
    \end{minipage}
      \\
    \vspace{-20em}
     \rotatebox[origin=Bc]{90}{\parbox{2.8cm}{{\scriptsize \ \ \ \ \ \ \ \ \  }}}
    &
      \begin{minipage}{.2\textwidth}
      \includegraphics[width=2.8cm,height=1.5cm, keepaspectratio]{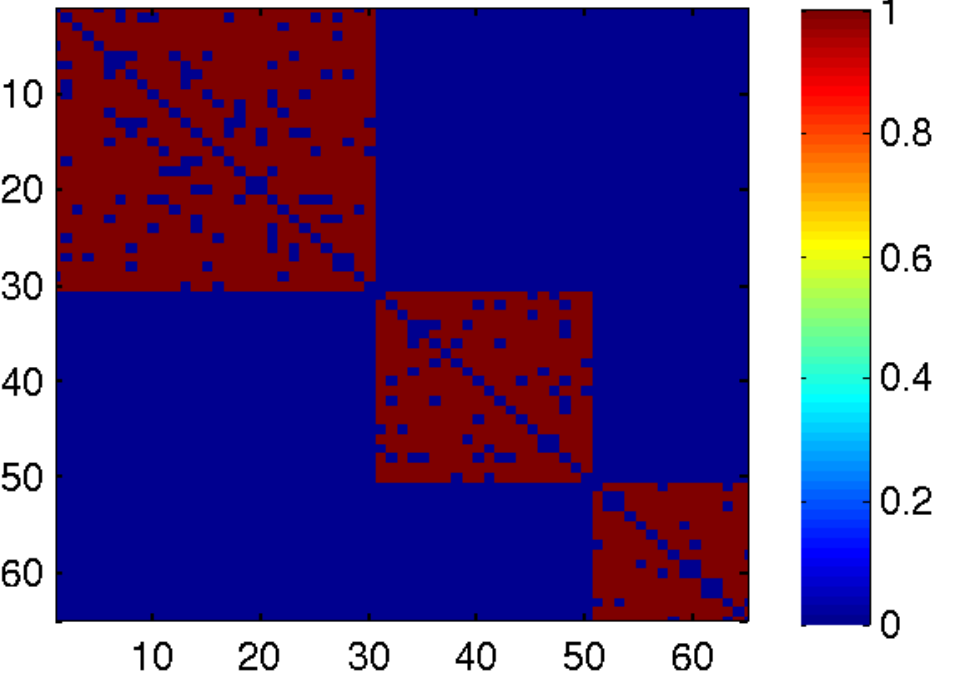}
    \end{minipage}
    &
    %\begin{minipage}[t]{5cm}
      \begin{minipage}{.21\textwidth}
      \includegraphics[width=2.8cm,height=1.5cm, keepaspectratio]{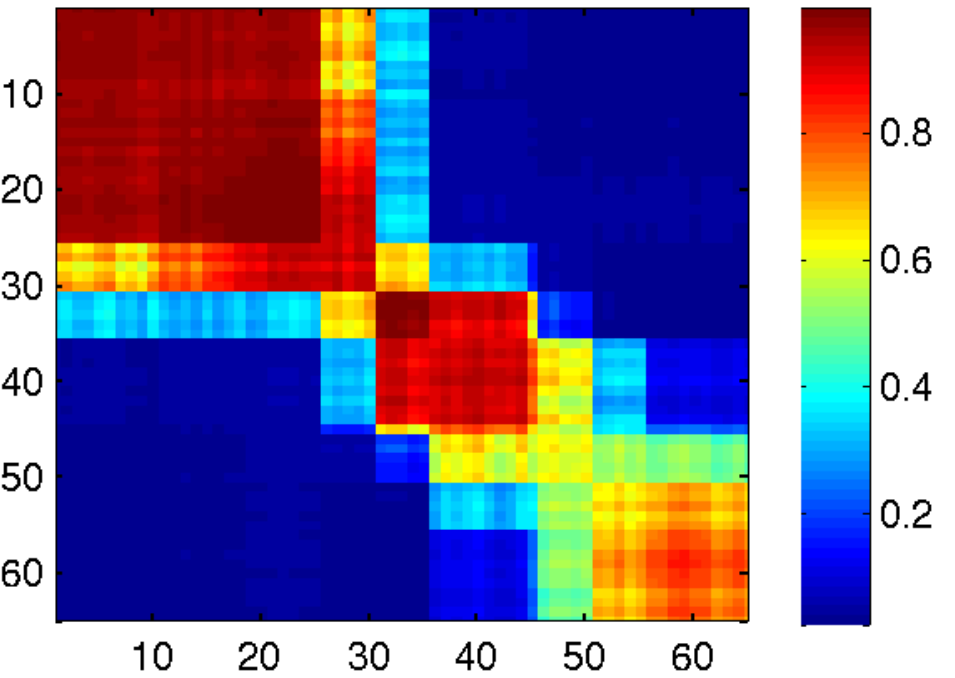}
    \end{minipage}
    %\end{minipage}
    & 
    %\begin{minipage}{5cm}
      \begin{minipage}{.21\textwidth}
      \includegraphics[width=2.8cm,height=1.5cm, keepaspectratio]{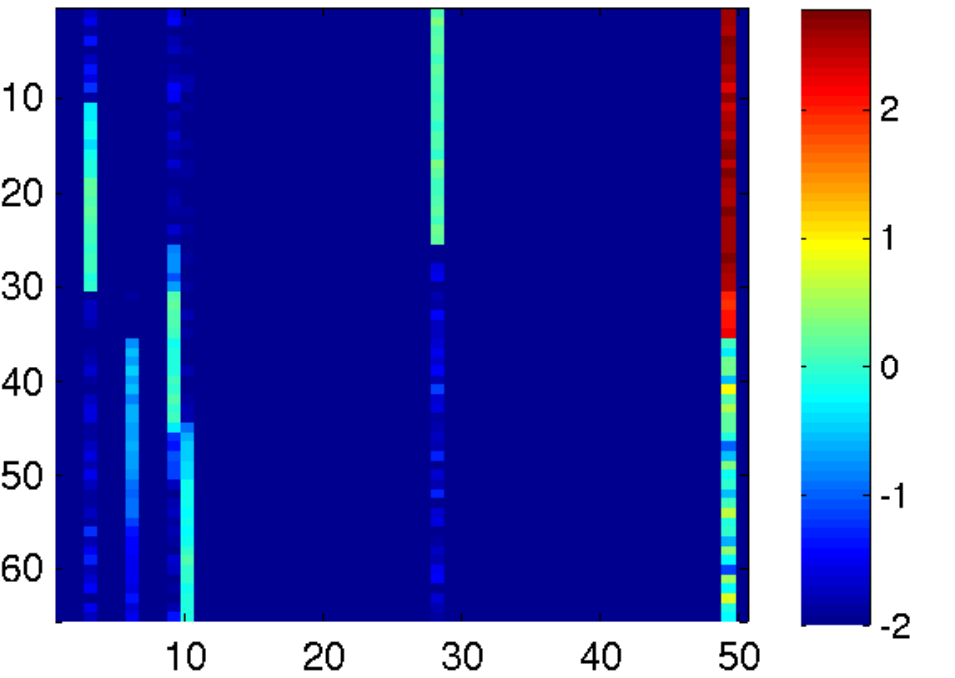}
    \end{minipage}
    %\end{minipage}
    &  
    %\begin{minipage}{5cm}
      \begin{minipage}{.21\textwidth}
      \includegraphics[width=2.8cm,height=1.5cm, keepaspectratio]{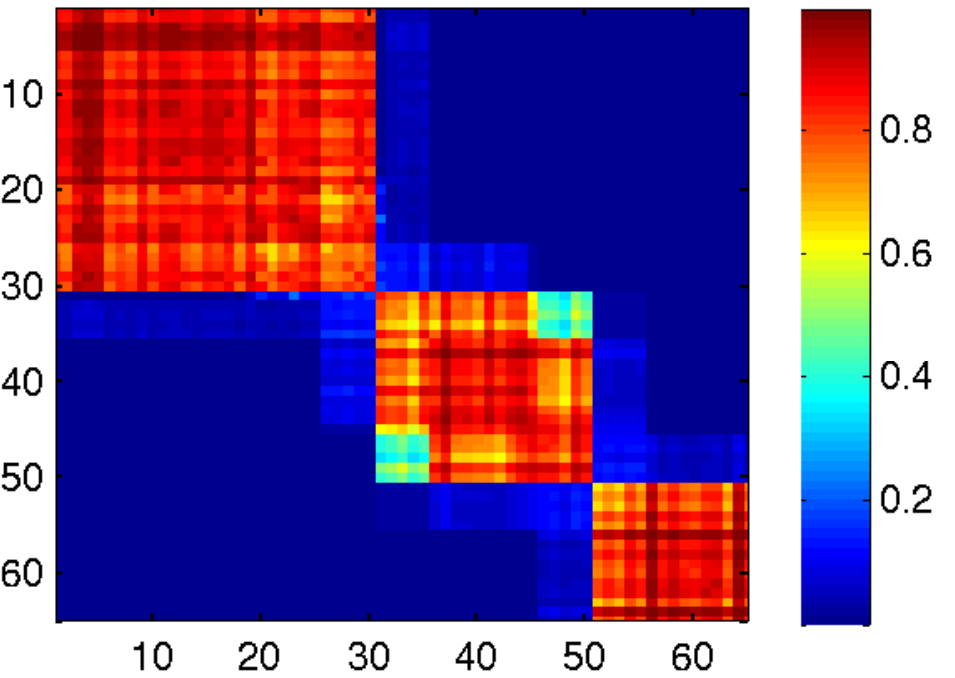}
    \end{minipage}
    &  
    %\begin{minipage}{5cm}
      \begin{minipage}{.21\textwidth}
      \includegraphics[width=2.8cm,height=1.5cm, keepaspectratio]{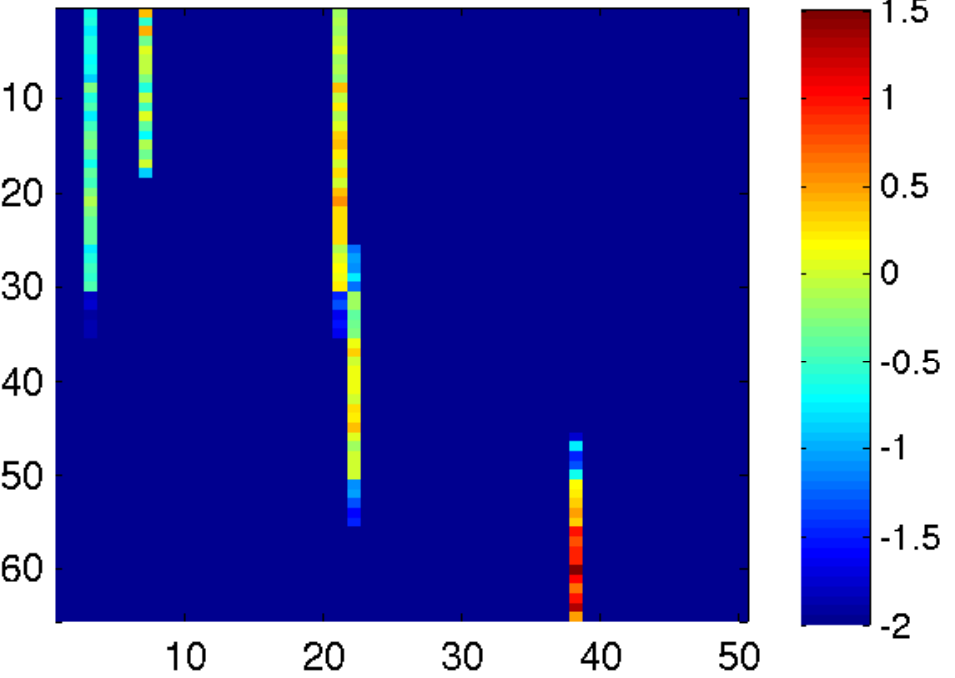}
    \end{minipage}\\
    & \ \ \ \ \ \ \ \ \ \  {\footnotesize (a)} & \ \ \ \ \ \ \ \ \ \ {\footnotesize (b)} & \ \ \ \ \ \ \ \ \ \ {\footnotesize (c)} & \ \ \ \ \ \ \ \ \ \ {\footnotesize (d)} & \ \ \ \ \ \ \ \ \ \  {\footnotesize (e)} \\ %\hline 
  \end{tabular}
  % \vspace{-1em}
  \caption{Dynamic community detection on synthetic data. We generate a dynamic network with five time snapshots as shown in column (a) ordered from top to bottom. The link probabilities estimated by D-GPPF and DPGM are shown in column (b) and (d). The association weights of each node to the latent groups can be calculated by $r_{tk}
\phi_{nk}$ for D-GPPF and $\phi_{nk}^{t}\lambda_{kk}$ for DPGM as shown in column (c) and (e), respectively. The pixel values are displayed on $\log_{10}$ scale.}
  \label{Synthetic}
%\end{table}
\end{figure*}
\begin{table*}[!htbp]
\small
\centering
\caption{\label{Prediction} {Missing links prediction. We highlight the performance of the best scoring model in bold.}}
%\vspace{-1em}
\begin{tabular}{|l|c|c|c|c|c|c|c|}\hline
& \multicolumn{2}{c|}{NIPS 70} & \multicolumn{2}{c|}{DBLP 96} & \multicolumn{2}{c|}{Enron 61}  \\ \hline %\cmidrule(r){2-7}%\cmidrule(l){8-15}
% Factors & 1 & 2 & 3 & 4 & 5 & 6 & 1 & 2 & 3 & 4 & 5 & 6 & 7 & 8\\ % 
Model & AUC-ROC & AUC-PR & AUC-ROC & AUC-PR & AUC-ROC & AUC-PR \\
\hline
HGP-EPM & ${0.947} \pm 0.003$ & ${0.146} \pm 0.030$ & $0.936 \pm 0.004$ & ${0.268} \pm 0.025$ & $0.931 \pm 0.003$ & $0.432 \pm 0.016$  \\\hline
D-GPPF &  $0.943 \pm 0.009$ & $0.072 \pm 0.023$ & $0.924 \pm 0.004$ & $0.249 \pm 0.010$ & $0.963 \pm 0.002$ & $0.649 \pm 0.022$ \\\hline
DRIFT & $0.963 \pm 0.002$ & $0.148 \pm 0.015$ & $0.965 \pm 0.001$ & ${0.335} \pm 0.001$ & $\mathbf{0.975} \pm 0.006$ & $\mathbf{0.736}\pm 0.030$ \\\hline
DPGM (batch) & $\mathbf{0.970} \pm 0.003$ & $\mathbf{0.169} \pm 0.028$ & $\mathbf{0.967} \pm 0.002$ & $\bm{0.339} \pm 0.006$ & $0.971\pm 0.002$ & $ 0.712 \pm 0.004$ \\\hline
% DPGM (online) & ${0.961} \pm 0.021$ & ${0.155} \pm 0.040$ & $0.957 \pm 0.003$ & ${0.293} \pm 0.026$ & $0.965 \pm 0.003$ & $0.689 \pm 0.015$  \\\hline
\hline
& \multicolumn{2}{c|}{NIPS 5K} & \multicolumn{2}{c|}{DBLP 7K} & \multicolumn{2}{c|}{Enron 2K}  \\ \hline %\cmidrule(r){2-7}%\cmidrule(l){8-15}
% Factors & 1 & 2 & 3 & 4 & 5 & 6 & 1 & 2 & 3 & 4 & 5 & 6 & 7 & 8\\ % 
Model & AUC-ROC & AUC-PR & AUC-ROC & AUC-PR & AUC-ROC & AUC-PR \\
\hline
HGP-EPM &  $0.928 \pm 0.010$ & ${0.121} \pm 0.008$ & $0.897 \pm 0.016$ & $0.055 \pm 0.008$ & $0.975 \pm 0.003$ & $0.324 \pm 0.006$  \\\hline
D-GPPF &  $0.928 \pm 0.006$ & ${0.119} \pm 0.006$ & $0.858 \pm 0.002$ & $0.054 \pm 0.003$ & $0.976 \pm 0.001$ & $0.305 \pm 0.003$  \\\hline
DPGM (batch) & $\mathbf{0.930} \pm 0.002$ & $\bm{0.129} \pm 0.004$ & $\mathbf{0.923} \pm 0.011$ & $\mathbf{0.057} \pm 0.001$ & $\mathbf{0.983} \pm 0.001$ & $\mathbf{0.398} \pm 0.006$ \\\hline
DPGM (online) & ${0.929} \pm 0.002$ & ${0.104} \pm 0.005$ & $0.911 \pm 0.001$ & $0.051 \pm 0.003$ & $0.976 \pm 0.001$ & $0.346 \pm 0.005$  \\\hline
\end{tabular}
\end{table*}
%\vspace{-0em}
\begin{table*}[!htbp]
\small
\centering
\caption{\label{TimeCost} {Comparison of per-iteration computation time (seconds).}}
%\vspace{-1.0em}
\begin{tabular}{|l|c|c|c||l|c|c|c|}\hline
 & {NIPS 70} & {DBLP 96} & {Enron 61} && {NIPS 5K} & {DBLP 7K} & {Enron 2K} \\ \hline %\cmidrule(r){2-7}%\cmidrule(l){8-15}
D-GPPF &  0.0388 & 0.1350 & 0.2161 & D-GPPF & 7.6440 & 7.9160 & 7.8227 \\\hline
DRIFT & 11.7047 & 42.1853 & 24.7505  & {DPGM (batch)} & 10.6240 & 15.6576 & 15.8584 \\\hline
 %\vspace{0.2em}
DPGM (batch) & 0.1283 & 0.6302 & 0.7364 & DPGM (online) & 8.9501 & 10.8152 & 10.4521 \\\hline
 % DPGM (online Gibbs) & 0.1050 & 0.3173 & 0.3970 \\\hline
\end{tabular}
\end{table*}

We adapt the synthetic example used in~\cite{DGPPF} to generate a dynamic network of size $65\time65$ that evolve over five time slices as shown in Fig.~\ref{Synthetic}.  More specifically, we generate three groups at $t=1$, and split the second group at $t = 2$. From $t=3$ to $4$, the second and third group merge into one group.
In Fig.~\ref{Synthetic}, column (b) and (d) show the discovered latent groups over time by D-GPPF and DPGM, respectively. D-GPPF captures the evolution of the discovered groups but has difficulties to characterize the changes of node-group memberships over time. Our model (DPGM) can detect the dynamic groups quite distinctively. We also show the associations of the nodes to the inferred latent groups by D-GPPF and DPGM in column (c) and (e), respectively. In particular, we calculate the association weights of each node to the latent groups as $r_{tk}\phi_{nk}$ for D-GPPF and $\phi_{nk}^{t}\lambda_{kk}$ for DPGM. For both models, most of the redundant groups can effectively be shrunk even though we initialize both algorithms with $K = 50$. The node-group association weights estimated by DPGM vary smoothly over time and capture the evolution of the node-group memberships, which is consistent to the ground truth shown in column (a). 
% We also found that the numbers of the inferred groups for both models seem to be greater than our expectation. This intuitively makes sense because the nonnegative memberships allow each node to be affiliated to multiple groups with different weights. Hence, a few redundant groups need to be generated to represent the overlapping parts among the dominating groups. 

\subsection{Missing link prediction}
For the task of missing link prediction, we randomly hold out $20\%$ of the observed interactions (either links or non-links) at each time as test data. The remaining data is used for training. HGP-EPM, DRIFT and D-GPPF are considered as the baseline methods. We train a HGP-EPM model on the training entries for each time slice separately. For each method, we use 2000 burn-in iterations, and collect 1000 samples of the model posterior. We estimate the posterior mean of the link probability for each held-out edge in the test data by averaging over the collected Gibbs samples. We then use these link probabilities to evaluate the predictive performance of each model by calculating the area under the curve of the receiver operating characteristic (AUC-ROC) and of the precision-recall (AUC-PR). Table~\ref{Prediction} shows the average evaluation metrics for each model over 10 runs. Overall, our model (DPGM) shows the best performance. 
We observe that both DRIFT and DPGM outperform D-GPPF because the evolution of individual node-group memberships are explicitly captured in these two models. D-GPPF essentially assumes that the nodes' memberships are static over time and thus has difficulties to fully capture the dynamics of each node's interactions caused by the same node's memberships' evolution. We see that DPGM outperforms its static counterpart, HGP-EPM, via capturing the evolution of nodes' memberships over time.
We also compare per-iteration computation time of each model (all models are implemented in Matlab), as shown in Table~\ref{TimeCost}. The computational cost of DRIFT scales quadratically with the number of nodes. Both D-GPPF and DPGM are much faster than DRIFT because the former two models scale only with the number of non-zero edges.
% More specifically, DPGM has $\mathcal{O}(N^{e} + NKT)$ computation, where $N^e$ is the number of non-zero edges.
We also report per-iteration computation time of GPPF and DPGM with Matlab/MEX/C implementation on medium-scale data in Table~\ref{TimeCost}.

%% CONCLUSION
\section{Conclusion}
We have presented a probabilistic model for learning from dynamic relational data. The evolution of the underlying structure is characterized by the Markovian construction of latent memberships. We also proposed efficient batch and online Gibbs algorithms that make use of the data augmentation technique. Experimental results on synthetic and real datasets illustrate our model's interpretable latent representations and competitive performance.
Our model is dedicated to dynamic networks modeling but can be considered for other related problems such as dynamic multi-relational graph model~\cite{NDKG}. 
Another interesting direction is to scale up the model inference algorithm via stochastic gradient variational Bayes~\cite{SGVB}.

\section{Acknowledgments}
The authors thank Adrian \v{S}o\v{s}i\'{c} and the reviewers for constructive comments. This work is funded by the European Union's Horizon 2020 research
and innovation programme under grant agreement 668858.
\bibliographystyle{aaai}
\bibliography{ref}

\end{document}